# Advancing C–C Coupling of Electrocatalytic $CO_2$ Reduction Reaction for $C_{2+}$ Products

*Guangyuan Liang,[1] Sheng Yang,[2] Chao Wu,[3] Yang Liu,[1] Yi Zhao,[1] Liang Huang,[4] Shaowei Zhang,[5] Shixue Dou,[3] Hongfang Du,[1]\* Dandan Cui,[6]\* Liangxu Lin[1]\**

[1]Strait Institute of Flexible Electronics (SIFE, Future Technologies), Fujian Key Laboratory of Flexible Electronics, Fujian Normal University and Strait Laboratory of Flexible Electronics (SLoFE), Fuzhou, 350117, P. R. China

[2]School of Energy Science and Engineering, Central South University, Changsha, 410083, P. R. China

[3]Institute of Energy Materials Science, University of Shanghai for Science and Technology, Shanghai, 200093, P. R. China

[4]The State Key Laboratory of Refractories and Metallurgy, Wuhan University of Science and Technology, Wuhan, 430081, P. R. China

[5]College of Engineering, Mathematics and Physical Sciences, University of Exeter, Exeter EX4 4QF, United Kingdom

[6]School of Physics, Beihang University, Beijing 100191, P. R. China

*Correspondence: ifelxlin@fjnu.edu.cn (L. Lin), ifehfdu@fjnu.edu.cn (H. Du), cuidan@buaa.edu.cn (D. Cui)

**ABSTRACT:** The production of multicarbon ($C_{2+}$) products through electrocatalytic $CO_2$ reduction reaction ($CO_2RR$) is crucial to addressing global environmental challenges and advancing sustainable energy solutions. However, efficiently producing these high-value chemicals via C–C coupling reactions is a significant challenge. This requires catalysts with optimized surface configurations and electronic properties capable of breaking the scaling relations among various intermediates. In this report, we introduce the fundamentals of electrocatalytic $CO_2RR$ and the mechanism of C–C coupling. We examine the effects of catalytic surface interactions with key intermediates and reaction pathways, and discuss emerging strategies for enhancing C–C coupling reactions toward $C_{2+}$ products. Despite varieties of these strategies, we summarize direct clues for the proper design of the catalyst for the electrocatalytic $CO_2RR$ towards $C_{2+}$ products, aiming to provide valuable insights to broad readers in the field.

## 1. Introduction

The electrocatalytic $CO_2$ reduction reaction ($CO_2RR$) represents a pivotal technology in tackling global environmental challenges and meeting the increasing demand for sustainable energy solutions.[1-5] Early investigations in the field have predominantly focused on the production of $C_1$ products, such as $CH_4$ and CO, although their conversions are usually accompanied with the competing hydrogen evolution reaction (HER).[6-8]

By comparison, multicarbon products ($C_{2+}$) offer greater commercial values and broader potentials owing to their high energy density and significant market demand.[9-13] However, the efficient production of $C_{2+}$ products via C–C coupling remains challenging. This difficulty arises largely from the need for a highly ordered atomic arrangement to stabilize coupling intermediates, and to overcome substantial activation barriers. Additionally, the hydrogenation of low-carbon intermediates is often kinetically more favorable than the C–C coupling.[11] The binding affinities of various intermediates on the catalyst also differ, leading to a range of C–C coupling products, where individual intermediate may yield different products through hydrogenation and reduction. In this context, the catalyst must disrupt the scaling relations among intermediates,[14-16] necessitating its precise design and fabrication.

To date, strategies such as facet and phase engineering,[17,18] strain and relaxed bonding structures,[19,20] hybrid structures and surface functionalization,[21-23] chemical doping and single atom catalysts (SACs),[24,25] and defect engineering[26,27] have been employed to optimize electrocatalytic $CO_2$RR (**Figure 1a**). Despite significant advances in producing $C_{2+}$ products, particularly with metallic copper-based catalysts,[28,29] current systems still fall short of commercial requirements (current density >200 mA cm$^{-2}$, Faradaic efficiency >70%, stability >100 hours). The performance of most catalysts is limited by scaling relations that govern the adsorption of various intermediates, making it challenging to achieve high efficiency and selectivity for $C_{2+}$ products.

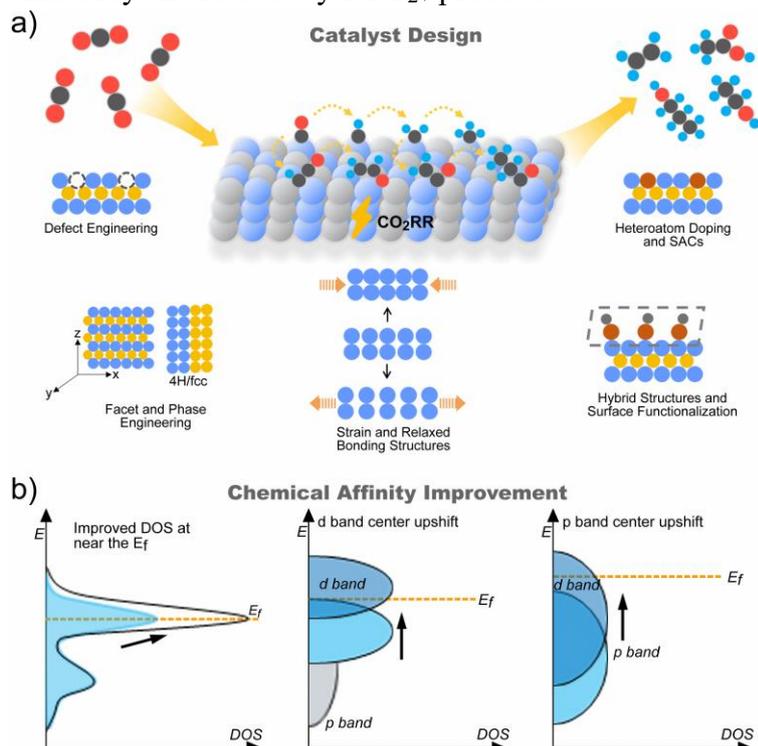

**Figure 1. a)** Various strategies in regulating the electrocatalytic $CO_2$RR. **b)** Diagrams show how material's chemical affinity can be improved by enhancing density of state (DOS) at near the Fermi level (left), upshifting d band (middle) and p band centers (right). Reproduced with permission.[38] Copyright 2024, American Chemical Society.

While most reports focus on engineering techniques to improve the efficiency of electrocatalytic $CO_2$RR,[30-33] few have examined the rational design for selectively producing $C_{2+}$ products. In this report, building on the fundamentals and reaction pathways involved in electrocatalytic $CO_2$RR, we discuss emerging strategies for the rational design of catalysts for $C_{2+}$ products. Focusing on the C–C coupling mechanisms of key intermediates and the primary factors influencing C–C bond formation, we analyze recent advances in the structure-property-product relationships of catalysts, aiming to provide valuable insights for developing novel catalysts for the selective and efficient electrocatalytic $CO_2RR$ of $C_{2+}$ products.

## 2. Fundamentals

Before discussing various catalytic pathways, we briefly introduce the fundamentals of electrocatalytic $CO_2RR$ and the concept on the regulation of binding affinity. A typical electrocatalytic $CO_2RR$ process consists of four key steps: (1) $CO_2$ adsorption, (2) $CO_2$ activation, (3) multiple electron/proton transfer processes, and (4) desorption of product.[34] Owing to the inherent stability of $CO_2$, the activation step requires considerable energy to convert the linear $CO_2$ molecule into a bent radical anion ($CO_2^{*-}$), which is the rate-determining step in electrocatalytic $CO_2RR$.[35] The reaction products are complex, as the small potential differences between the coupling of the highly reactive $CO_2^{*-}$ radical anion and various intermediates result in multiple possible outcomes. Depending on the number of transferred electrons and protons, the electrocatalytic $CO_2RR$ yields approximately 16 products (**Table 1**).[36,37]

**Table 1.** Main products involved in the half reaction process of electrocatalytic $CO_2RR$ (25 °C, pH = 7, 1 atmosphere of gases), and their corresponding standard redox potential (versus reversible hydrogen electrode potential/RHE, $V_{RHE}$).[36,37]

| Products | n | Reaction pathway | E ($V_{RHE}$) |
|---|---|---|---|
| $H_2$ | 2 | $2H^+ + 2e^- \rightarrow H_2$ | 0 |
| CO | 2 | $CO_2 + 2H^+ + 2e^- \rightarrow CO + H_2O$ | -0.1 |
| HCOOH | 2 | $CO_2 + 2H^+ + 2e^- \rightarrow HCOOH$ | -0.19 |
| Others | 4 | $CO_2 + 4H^+ + 4e^- \rightarrow HCHO + H_2O$ | -0.09 |
| | 6 | $CO_2 + 6H^+ + 6e^- \rightarrow CH_3OH + H_2O$ | +0.03 |
| | 8 | $CO_2 + 8H^+ + 8e^- \rightarrow CH_4 + 2H_2O$ | +0.18 |
| | 8 | $2CO_2 + 8H^+ + 8e^- \rightarrow CH_3COOH + 2H_2O$ | +0.13 |
| | 10 | $2CO_2 + 10H^+ + 10e^- \rightarrow CH_3CHO + 3H_2O$ | +0.06 |
| | 12 | $2CO_2 + 12H^+ + 12e^- \rightarrow C_2H_4 + 4H_2O$ | +0.08 |
| | 12 | $2CO_2 + 12H^+ + 12e^- \rightarrow C_2H_5OH + 3H_2O$ | +0.09 |
| | 14 | $2CO_2 + 14H^+ + 14e^- \rightarrow C_2H_6 + 4H_2O$ | +0.15 |
| | 18 | $3CO_2 + 18H^+ + 18e^- \rightarrow C_3H_7OH + 5H_2O$ | +0.11 |

Understanding the reaction pathways from atomic and molecular perspectives is essential for elucidating the interactions between reactants and catalytic sites, as well as the transformation of surface-bound intermediates. The advent of in situ techniques, combined with theoretical calculations and transport behavior models, has enabled the identification of plausible reaction pathways for $C_{2+}$ product formation. This integration has enriched our understanding on the interactions among intermediates, reactants, and catalytic sites, along with the structure-property-product relationship. These insights facilitate the rational design of electrocatalytic systems. In the following section, we discuss the primary products of electrocatalytic $CO_2RR$, focusing particularly on the pathways for $C_{2+}$ products and identifying key factors that promote C–C coupling reactions. These strategies are mainly focused on the regulation of electronic structure and adsorption configurations. In most cases, an improved binding affinity of a material for $CO_2RR$ intermediates can be attributed to an increased density of dangling bond states near the Fermi level and an upshift of the d/p band center (**Figure 1b**). These changes elevate the anti-bonding states above the Fermi level, thereby increasing the availability of electronic states in the bonding orbitals.[38] As a result, the bonding level of the catalytic site is lowered, facilitating more effective hybridization with the s/p orbitals of the $CO_2RR$ intermediates.

## 3. Main Pathways of the Electrocatalytic $CO_2RR$

### 3.1. CO and HCOOH

$HCOO^-$ or HCOOH may form through an intermediate that binds to transition metal catalyst electrode either by one oxygen atom (monodentate) or two oxygen atoms (bidentate) (**Figure 2a**, and the top pathway in **Figure 2c**).[39-41] The intermediate can be generated by reacting with *H via $CO_2$ insertion into the metal-hydrogen bond, or by direct protonation with $H^+$ from the solution. An alternative pathway is involving the $CO_2^{*-}$ radical reaction with the adjacent water molecule or proton to yield $HCOO^-$ or HCOOH (middle pathway in **Figure 2c**).[41,42] Nevertheless, this pathway is unlikely to be proceed owing to the very negative redox potential (-1.48 $V_{RHE}$, pH=7) of the direct formation of $CO_2^{*-}$ by single electron transfer.[43] Notably, the presence of $HCO_3^-$ can enhance the production of $HCOO^-$. $SnO_2$, for example, is initially reduced to Sn (II) oxyhydroxide, and then reacts with $CO_2$ to form a surface-bound carbonate.[44] The carbonate undergoes a transformation of two electrons and a proton to form $HCOO^-$. The Sn (II) oxyhydroxide is then recovered upon the desorption of $HCOO^-$ (**Figure 2b**).

Conversely, the electrocatalytic $CO_2RR$ to CO is a straightforward two-electron transfer process (bottom pathway in **Figure 2c**).[45] In this process, the adsorbed $CO_2$ molecule is reduced to *COOH via a concerted proton-electron reaction ($H^+ + e^-$). The second proton-electron pair can subsequently react with *COOH to produce $H_2O$ and CO, which then desorbs from the electrode in gas form. The initial reduction of $CO_2$ to *COOH is inhibited by weak binding of *COOH, while the final step of CO desorption is hindered by strong binding of *CO. These two steps are rate-limiting reactions, as indicated by density functional theory (DFT) calculations.[46]

An alternative pathway to *COOH involves decoupled electron and proton transfers, beginning with the formation of the $CO_2^{*-}$ radical anion (middle pathway in **Figure 2c**). However, as mentioned earlier, the initial activation of $CO_2$ through single electron transfer in this step is unfavorable owing to the very negative redox potential.[43] To generate CO and $HCOO^-$/HCOOH, two proton-electron pairs must be transferred to the adsorbate, albeit through different intermediates. Nørskov et al. have identified two distinct adsorption configurations based on thermodynamic calculations: *OCHO (with two oxygen atoms as binding sites) and *COOH (with one carbon atom as a binding site).[47] It is suggested that *OCHO primarily converts to HCOOH, while *COOH converts to CO.[47] These two catalytic pathways align well with the routes described above in **Figure 2a** and **Figure 2c** (bottom), respectively. In this context, the binding energies of *OCHO and *COOH on the catalyst serve as descriptors for the products HCOOH and CO, respectively, exhibiting clear volcano trends across a range of transition metals.[47]

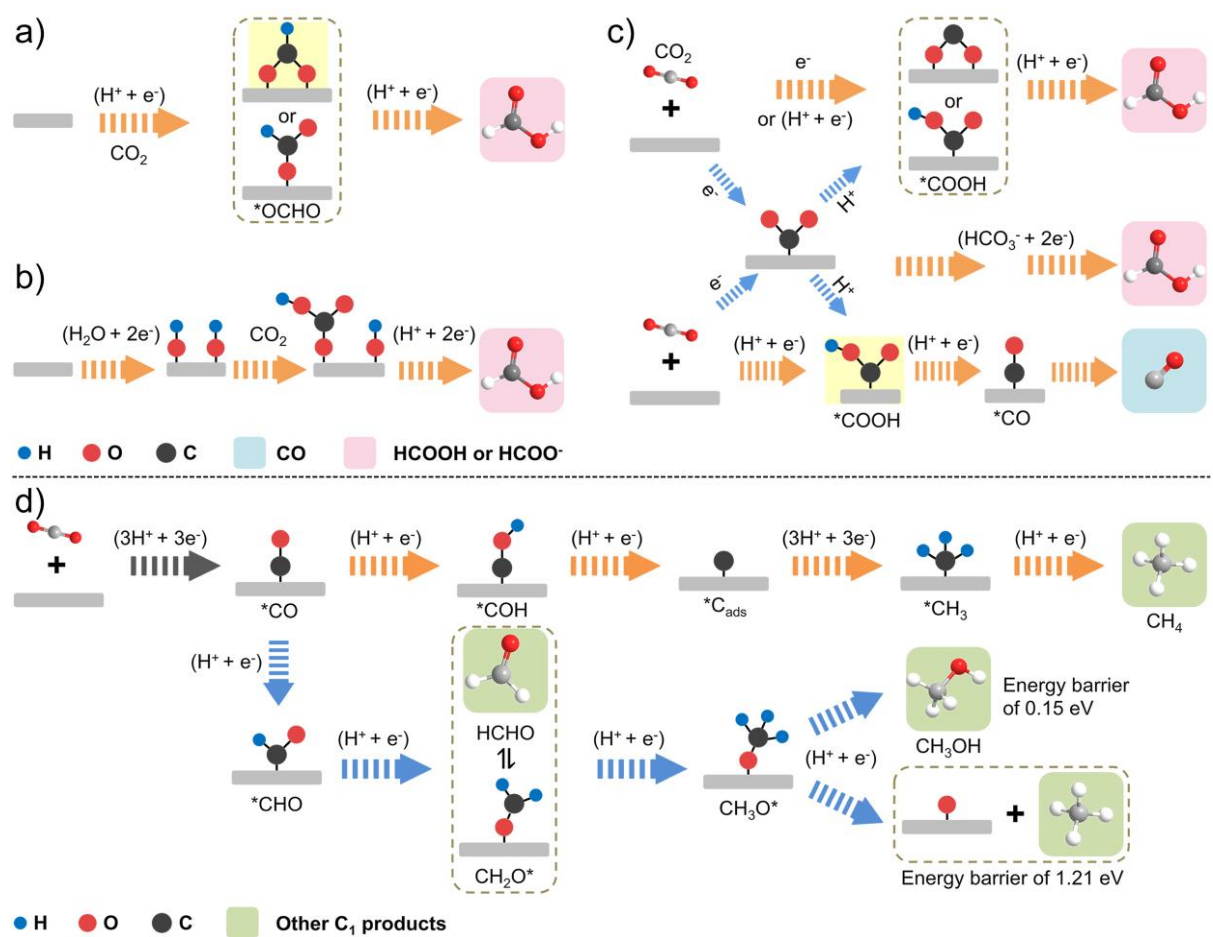

**Figure 2.** Possible electrocatalytic $CO_2$RR pathways to HCOOH, CO and other $C_1$ products. **a)** HCOOH generation via monodentate or bidentate intermediates.[41] **b)** HCOOH generation via surface-bound carbonate intermediates.[44] **c)** HCOOH and CO generation via $CO_2$*[-] radical intermediates.[40-42,45] The catalytic routes, with yellow pattern highlighted intermediates, match well with that described by Nørskov et al.[47] **d)** Possible electrocatalytic $CO_2$RR pathways to HCHO, $CH_3OH$, and $CH_4$.[48,49] These figures are prepared following above listed literatures.

## 3.2. Other hydrocarbons and oxygenated compounds
### 3.2.1. $C_1$ products
To produce HCHO, $CH_3OH$, and $CH_4$, *CO is likely a common intermediate.[41] DFT calculations indicate that the most thermodynamically favorable pathway begins with the formation of *CO, followed by hydrogenation to *CHO, $CH_2O$* (which desorbs to form HCHO), and $CH_3O$* (as illustrated by the blue arrows in **Figure 2d**).[48] The $CH_3O$* intermediate can then be reduced to $CH_3OH$ or converted to a complex of $CH_4$ and *O. However, the formation of $CH_4$ through C–H bonding and C–O dissociation is calculated to be kinetically prohibitive, with an energy barrier of 1.21 eV, which is significantly higher than the 0.15 eV barrier for $CH_3OH$ production (**Figure 2d**).[49]

To produce $CH_4$, a possible pathway involves the reduction of *COH intermediates to adsorbed *C ($C_{ads}$, as the orange arrows in **Figure 2d** indicates).[49] This surface carbon can be further reduced to *CH, *$CH_2$, *$CH_3$, and ultimately to $CH_4$, examined using in situ X-ray photoelectron spectroscopy (XPS) and Auger electron spectroscopy (AES).[50]

### 3.2.2. $C_2$ products

The formation of $C_2$ products, such as $C_2H_4$, $CH_3CHO$, and $C_2H_5OH$, is a complex process with multiple possible pathways, which remain the subject of ongoing debate. Compared to $C_1$ products, the formation of $C_{2+}$ products requires a higher binding affinity to stabilize the *CO intermediate and more catalytic active sites to facilitate the C–C coupling (will be further discussed in **Section 3.3**).

In the "carbene" mechanism of electrocatalytic $CO_2RR$ for $C_{2+}$ products, $C_2H_4$ is formed either through the coupling of two *$CH_2$ species[49] or via CO insertion in a Fischer-Tropsch-like step (**Figure 3a**),[51] where the *$COCH_2$ intermediate further undergoes hydrogenation to yield acetaldehyde and ethanol. Alternatively, the key intermediate *$C_2O_2$ can be generated through the dimerization of *CO, which is then protonated to form the $CH_2CHO$* intermediate (**Figure 3b**).[52,53] The $CH_2CHO$*, a key shared intermediate in the formation of $C_2H_4$ and $C_2H_5OH$, can be further reduced to $C_2H_4$ or $CH_3CHO$, which can subsequently produce $C_2H_5OH$, with $C_2H_4$ being the kinetically favored product on Cu surfaces.[54]

Additionally, the formation of other $C_2$ products, such as $CH_3COOH$ and $CH_3COO^-$, is proposed to involve a nucleophilic attack by adsorbed $CO_2$*$^-$ on the reduced *$CH_3$ species (**Figure 3c**).[55,56] In a separate study with integrated electrokinetic data and in situ infrared spectroscopy, it was observed that two adsorbed $CO_2$*$^-$ can couple to form *COOCOO, which may serve as a crucial intermediate for the formation of $CH_3COOH$.[25]

### 3.2.3. $C_3$ products

Unlike the aforementioned $C_1$ and $C_2$ products, the electrocatalytic $CO_2RR$ to long-chain products ($C_{3+}$) is a significant challenge because of the need for multiple C–C coupling steps. The formation of a *COCOCO intermediate, achieved through consecutive coupling of *CO key intermediates, is subsequently protonated and reduced to yield $C_3H_7OH$ (**Figure 3d**).[58]

Alternatively, the *CO intermediate can be initially reduced to the *$CH_2$ intermediate, which is then coupled and desorbed to produce $C_3H_6$ (**Figure 3d**).[59] Furthermore, *$CH_2$ can continue to undergo insertion reactions with intermediates (such as CO) to construct longer carbon chains and generate desired products (**Figure 3e**).[51] Understanding on the electrocatalytic $CO_2RR$ for $C_{3+}$ products is not so well established, compared to above $C_1$ and $C_2$ products, which will be not further extended in this discussion. Correspondingly, in the next discussions on associated strategies to improve the C–C coupling, the majority of cases are still the electrocatalytic $CO_2RR$ for $C_2$ products.

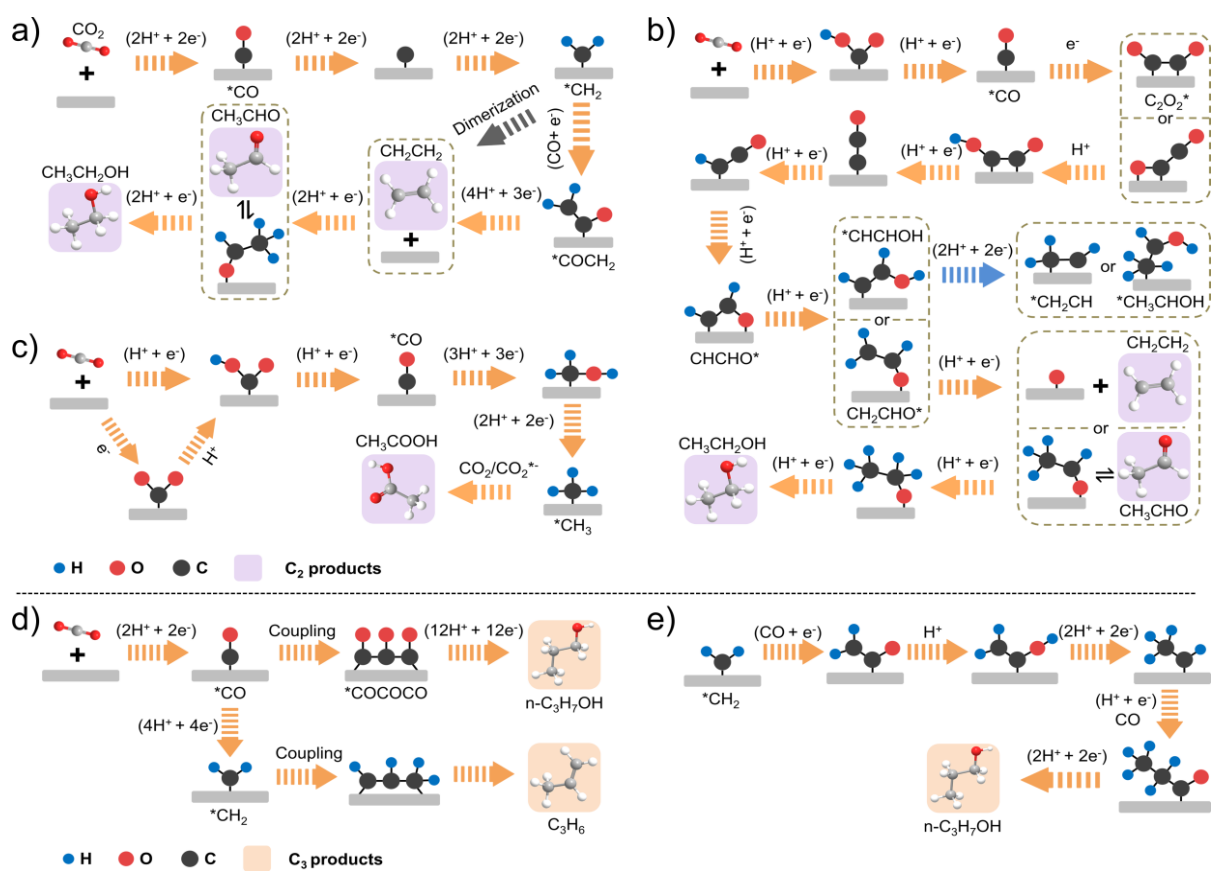

**Figure 3.** Possible electrocatalytic CO$_2$RR pathways to C$_2$ and C$_3$ products. **a)** Coupling of two *CH$_2$ species or CO insertion in a Fischer-Tropsch-like step.[51] **b)** *CO dimerization and bifurcated pathways to ethylene (*CH$_2$CH) and ethanol (*CH$_3$CHOH).[48,52,57] **c)** The production of CH$_3$COOH through the nucleophilic attack of CH$_3$ species by CO$_2$/CO$_2$*$^-$.[51] **d)** The production of C$_3$H$_6$ and n-C$_3$H$_7$OH though multistep coupling pathways.[58,59] **e)** CO insertion in a Fischer-Tropsch-like step.[51] These figures are prepared following above listed literatures.

### 3.3. Key factors of C–C couplings

Considering above various possible pathways, the *CO intermediate is often used to evaluate the selectivity for deep reduction products in the electrocatalytic CO$_2$RR. At the catalyst surface, if the *CO adsorption affinity is insufficient or the surface fails to stabilize *CO, CO is likely the primary product (**Figure 4a**). Conversely, for catalysts strongly bind with *CO, the catalyst surface may become poisoned and catalytically inactive (**Figure 4a**).

Catalysts such as copper, with a moderate binding affinity to *CO, can efficiently reduce *CO for further reactions.[60] Studies indicate that *CO can adopt diverse binding configurations on the catalyst, primarily including *CO$_{top}$ (where *CO is bonded to the top of a Cu atom through its C atom), *CO$_{bridge}$ (where the C atom in *CO is bonded to two Cu atoms), and *CO$_{hollow}$ configurations (where CO is adsorbed on terrace and defect sites) (**Figure 4a**).[61,62] As the surface coverage of *CO increases, the adsorption mode shifts from *CO$_{bridge}$ to *CO$_{top}$ (**Figure 4a**), where the coupling energy barrier between two *CO$_{top}$ species is minimized.[63,64] Therefore, a moderate *CO coverage on the catalyst can lower the energy barrier for C–C coupling while inhibiting the HER, thereby enhancing the conversion efficiency of C$_{2+}$ products (**Figure 4a**).

Variations in the adsorption strengths of different atoms on the catalyst can lead to significant changes of catalytic selectivity (**Figure 4b,c**). In the competitive reaction between CO and HCOOH, when the objective is CO production, $CO_2$ can only adsorb on the catalyst through its carbon atoms (**Figure 2c**). By modulating the binding energies of adsorbed carbon and oxygen, one can favorably or unfavorably influence the cleavage of C–O bonds, thereby designing the catalyst to enhance selectivity for specific products (**Figure 4b**). For example, the branching pathways for the intermediate CHCHO* (**Figure 3b**), in the production of ethylene and ethanol, are determined by the catalyst's affinity for C and O.[15] If the catalyst has stable binding energy to the carbon atom, subsequent protonation occurs preferentially on the O atom, and ethylene is more likely to be the primary product. Conversely, if protons preferentially attack the carbon atom in C=C bonds because of the catalyst's stable binding affinity to oxygen, the formation of the $CH_3CHO^*$ intermediate will be favored, leading to an increased production of ethanol over ethylene (the route follows the orange arrows in **Figure 3b**). In this scenario, further enhancement of binding affinity is not conducive to the selective production of ethanol, as the dissociation of C–O in the $*CH_3CHO$ intermediate would yield ethylene (the route follows the orange arrows in **Figure 3b**).

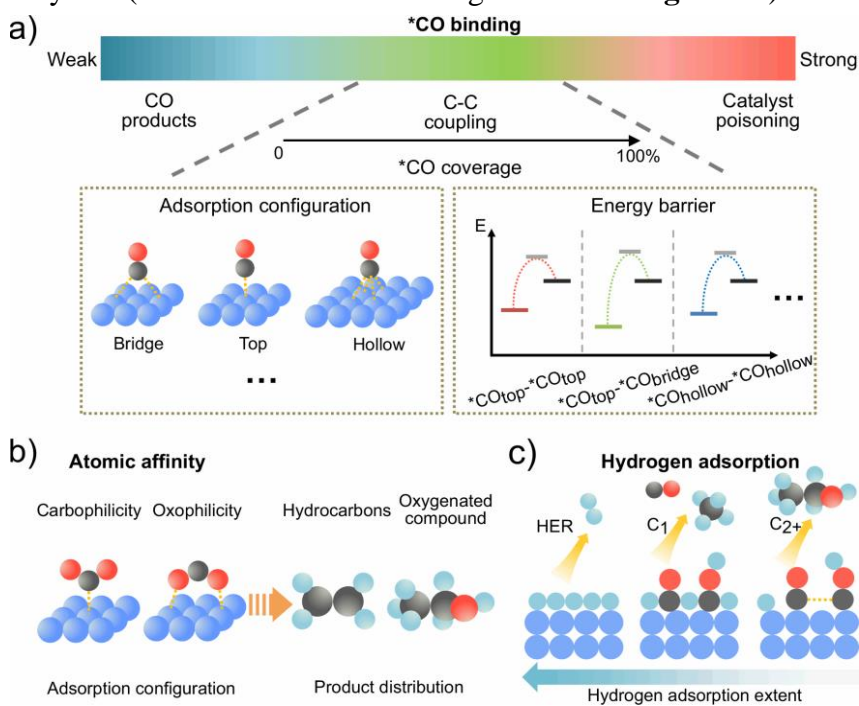

**Figure 4**. **a**) The effect of *CO binding strength and *CO coverage on the C–C coupling energy barrier at catalyst surface. **b**) The effect of oxophilicity or carbophilicity on the adsorption configuration of intermediates and product distribution. **c**) The degree of hydrogen adsorption affects intermediate evolution and target products.

In the context of deep reduction to $C_1$ products such as $CH_4$ and $CH_3OH$, multi-step hydrogenation of *CO is crucial. Moreover, the intermediates *CHO and *COH involved in the hydrogenation of *CO are also key players in the formation of $C_{2+}$ products. Because of the lower formation energies of C–H and H–H bonds,[65] and the kinetically more favorable hydrogenation of adsorbed $C_1$ intermediates compared to C–C couplings,[66,67] the C–C coupling emerges as the rate-determining step for $C_{2+}$ production. Therefore, inhibiting the further hydrogenation of $C_1$ intermediates and minimizing excessive hydrogen adsorption on the catalyst favorably promotes the C–C coupling (**Figure 4c**).

Furthermore, the electrocatalytic CO$_2$RR to C$_{2+}$ products is significantly influenced by reaction conditions, including local pH and applied overpotential.[13,68] As a result, researchers have focused on enhancing selectivity by optimizing macroscopic conditions, such as ion exchange membranes,[69-71] electrolytes,[72-74] and electrolytic cells.[75-77] However, to gain a deeper understanding of the catalytic mechanisms underlying C$_{2+}$ product formation, it is essential to undertake a precise design of the catalyst's structure, in conjunction with advanced characterization techniques.

## 4. Strategies for C$_{2+}$ products

To achieve low overpotentials, high current densities, high Faradaic efficiencies, and long-term stability of the electrocatalytic CO$_2$RR, various engineering strategies have been attempted. These strategies encompass defect engineering, heteroatom doping and SACs, facet and phase engineering, strain and relaxed bonding structures, hybrid structures and surface functionalization. This section provides a comprehensive summary and critical discussion of both theoretical and experimental studies related to these approaches, elucidating their contributions to the optimization of CO$_2$RR catalysts for C$_{2+}$ product formation.

### 4.1. Defect engineering

The term "defect engineering" is widely used in the literature but can be somewhat arbitrary when describing changes to the electronic structures and catalytic active sites of materials. Compared to ideal highly crystalline phases, defects may include vacancies, grain boundaries, structural transformations, strained systems, and relaxed bonding structures.[38,78] Relaxed bonding structures, often found in smaller materials, can be characterized as unsaturated binding systems with shifted d/p band centers. For instance, in carbon nanoribbons, increased C–C/C=C bond lengths enhance reactivity toward external molecules.[79] Strained systems and relaxed bonding structures are sometimes distinguished from traditional defect structures, a topic that will be further explored in the **Section 4.4**. Other defects can create structural asymmetries that induce electronic polarization in local defect structures.[38,80] In these contexts, both defect sites and neighboring atoms can act as catalytic active sites, depending on the nature of the electronic polarization and the reacting species.[81-83] In the case of CO$_2$RR, defects can significantly influence the electronic properties of the catalyst surface and its catalytic performance.[58,84-91] Studies have shown that defects can modify the energy barriers of rate-limiting steps, facilitating the adsorption, concentration, and confinement of reaction intermediates (e.g., *CO), ultimately promoting C–C coupling between two C$_1$ intermediates.[58,88-91] This section focuses on common defects, particularly vacancies and grain boundaries, to provide a general understanding of how catalyst electronic structures can be regulated.

Among various defects, vacancies represent the simplest and most abundant defect structures. Peng et al. investigated cubic copper nitride with varying N vacancies (Cu$_3$Nx) for electrocatalytic CO$_2$RR (**Figure 5a**).[92] These N vacancies were created using an

9 / 39

electrochemical lithium tuning strategy. DFT studies indicated that the instability of the OCCO* intermediate on the pristine $Cu_3N$ crystal surface hinders CO–CO coupling. However, this scenario changes with the formation of N vacancies. When the N vacancy density reaches 37.5%, the distance between copper sites approximates that of pure copper, resulting in increased electron cloud density that enhances adsorption of *CO. The optimized binding energy for CO–CO coupling is -0.142 eV, with a corresponding energy barrier of 0.366 eV. Consequently, the N-deficient $Cu_3N$ demonstrated outstanding $CO_2$-to-$C_2$ conversion efficiency, achieving a Faradaic efficiency (FE) of over 80% and a current density of approximately 300 mA cm$^{-2}$. Traditionally, defect structures are perceived as unstable owing to newly formed dangling bond states.[38] However, the N-deficient $Cu_3N$ was reported to exhibit high stability during electrocatalytic $CO_2RR$ at elevated current densities. These defective structures may reorganize or passivate to maintain electrochemical stability, similar to the behavior observed in vacancy-containing metal sulfide monolayers.[38,93-95]

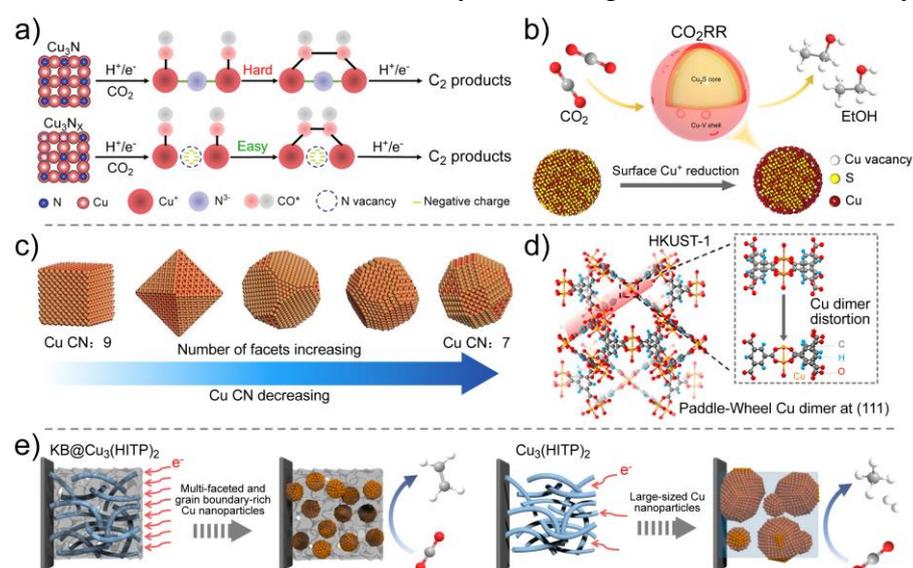

**Figure 5**. **a)** The electrocatalytic $CO_2RR$ on $Cu_3N$ and $Cu_3N_X$. Reproduced with permission.[92] Copyright 2021, Wiley. **b)** Schematics of $Cu_2S$-Cu-V design for the production of multi-carbon alcohols. Reproduced with permission.[96] Copyright 2018, Springer Nature. **c)** Schematics of Cu catalysts with different number of facets. CN: the Cu coordination number. Reproduced with permission.[98] Copyright 2024, Wiley. **d)** Schematics of the symmetric Cu dimer building block of HKUST-1 distorted to an asymmetric motif to form Cu clusters with a low coordination number. Reproduced with permission.[99] Copyright 2018, American Chemical Society. **e)** The regulation of $CO_2RR$ selectivity by $Cu_3(HITP)_2$-derived $Cu^0$ entities in the presence (left) or absence (right) of carbon support. Reproduced with permission.[57] Copyright 2021, Springer Nature.

In a related study, Cu vacancies were introduced into a core-shell structure, consisting of a core of $Cu_2S$ and a shell of copper with vacancies (**Figure 5b**).[96] Compared to pure Cu, the copper shell with vacancies exhibits an increased energy barrier for the conversion of $CO_2$ to ethylene, while the production of ethanol remains unaffected.[96] This phenomenon arises because the presence of Cu vacancies elongates the Cu–Cu bond length, facilitating binding with oxygen in the $CH_2CHO*$ intermediate at a bridge site, characterized by two O–Cu coordination numbers.[96] This configuration optimizes binding affinity and promotes ethanol formation. In contrast, in pure copper, the oxygen in the $CH_2CHO*$ intermediate binds through four O–Cu coordination numbers at a hollow position, resulting in excessive binding affinity that favors the dissociation of the C–O bond in $CH_2CHO*$, thereby facilitating ethylene production (see **Figure 3b**).[96]

These examples illustrate that vacancy formation in fully coordinated metal compounds (e.g., N vacancies in $Cu_3N$) can enhance the binding affinity for *CO, stabilizing the C–C coupling—a crucial step in the electrocatalytic $CO_2RR$ to $C_{2+}$ products (**Figure 3**). This phenomenon is consistent with the fundamentals of electrocatalysis and has been observed in other materials, such as $Cu_2O$.[97] However, in pure metals like copper, the binding affinity for intermediates is often too strong, necessitating a reduction in affinity to favor ethanol production, which competes with ethylene formation, as discussed above,[96] or the generation of longer-chain products.[97] In this context, the introduction of vacancies can lower the binding affinity (by reducing the coordination number with O-containing species) through modifications in the binding mode.

These concepts were further explored by Fang et al., who calculated the binding energy of *CO on various Cu-based catalysts with differing coordination numbers (**Figure 5c**).[98] As the coordination number of Cu sites decreased from 9 to 7, the binding affinity for *CO increased, resulting in a reduced energy barrier for C–C coupling. By fabricating various $Cu_2O$ polyhedra with controlled coordination numbers through an increased number of facets (**Figure 5c**), they demonstrated that $Cu_2O$ with the largest number of facets exhibited exceptional activity for ethylene production, achieving a FE of 72% at a current density of 800 mA $cm^{-2}$. In membrane electrode assembly cells, this catalyst maintained impressive performance over a 230-hour operation at a cell voltage of 3.5 V, with only marginal reductions in FE. This finding underscores that the binding affinity of $Cu_2O$ for intermediates is sufficiently high to favor ethylene formation. A similar strategy was reported by Nam et al., who distorted a Cu dimer precursor to achieve a low coordination number in the resulting Cu cluster product (**Figure 5d**).[99] Lin et al. also discussed the use of metastable or distorted local structures to enhance binding affinity through increased dangling bond states.[38] This improved the binding affinity of the Cu cluster, favoring ethylene formation (with FE for ethylene production increasing from 10% to 45%).[99]

Similarly, other defect engineering techniques can enhance the $CO_2RR$ by inducing electronic polarization to improve the binding affinity to *CO. For instance, grain boundaries have been introduced to increase the binding affinity of *CO in Cu-based metal-organic frameworks (MOFs) (**Figure 5e**).[57] In this context, the energy barriers for proton-coupled electron transfer to form the *CHCHOH intermediate—leading to the production of *$CH_2CH$ (ultimately yielding ethylene) and *$CH_3CHOH$ (ultimately yielding ethanol)—are -0.19 eV and 0.26 eV, respectively (following the blue arrows in **Figure 3b**). As a result, a stable FE of 60-80% for ethylene production is achieved over a wide potential range from -1.1 to -1.5 $V_{RHE}$. Gong et al. also demonstrated a high FE of 73% for $C_{2+}$ products, including ethylene, ethanol, and propanol, by creating grain boundaries in Cu catalysts.[100] The introduction of grain boundaries leads to under-coordinated surface sites and relatively extended Cu–Cu bonds, resulting in enhanced binding affinity.[100] This favors the stabilization of bidentate adsorbates (*COOH and *CHO) on the catalyst and lowers the energy barrier for CO–CHO coupling, rendering it both thermodynamically and kinetically favorable.[100] As illustrated in **Figure 3b**, the current consensus is that C–C coupling on Cu surfaces primarily occurs via the dimerization of two *CO molecules, followed by protonation to form a *COCOH or *COCHO intermediate.

Therefore, in most cases of defect engineering, a low coordination number (as indicated by the d band upshift in **Figure 1b**) and enlarged Cu–Cu bonds are considered advantageous for enhancing the binding affinity to CO, thereby facilitating selective C–C coupling. In comparison to $C_2H_4$, the production of $CH_3CH_2OH$ requires a slightly lower binding affinity of the catalyst to the $CH_2CHO$ intermediate through the O atom, which helps to prevent the dissociation of the C–O bond in $CH_2CHO*$ (**Figure 3b**). It is crucial to thoroughly consider the adsorption configurations of intermediates on the catalysts, as these configurations significantly influence the actual coordination environment. By employing appropriate DFT modeling to explore the most thermodynamically and kinetically favorable catalytic pathways, selective $C_{2+}$ production becomes attainable. This understanding serves as a foundation for optimizing catalysts for $C_{2+}$ production and is also applicable to other engineering techniques.

## 4.2. Chemical doping and SACs

Chemical doping involves the intentional introduction of heteroatoms into a host material to modify its electronic structure.[101,102] This process induces electronic polarization and charge redistribution, thereby altering the chemical affinity at the doping site and in neighboring regions to enhance catalytic activity.[78,80,82,83] The effects of electronic polarization and charge redistribution can be understood through electron shifts between atoms with differing electronegativities. Additionally, dopants with a larger atomic radius can cause geometric distortions in the host lattice, leading to the formation of new dangling bond states.[78,80] To date, a variety of metal and non-metal heteroatoms have been applied as dopants to optimize the electrocatalytic $CO_2RR$.[103-111]

Metal heteroatoms inherently possess a degree of catalytic activity for the electrocatalytic $CO_2RR$. These metal dopants not only modulate the electronic structure of active sites but also directly participate in the reaction, creating a synergistic catalytic effect. For example, Al dopants in $Cu(111)/Cu_2O(100)$ enhance the electronic density of Cu 3d orbitals near the Fermi level (see **Figure 1b**), increasing the chemical affinity for *CO intermediates and improving *CO coverage.[103] This modification effectively reduces the energy barrier for *CO−*COH coupling, thereby facilitating $C_{2+}$ products formation. In the subsequent bifurcation pathway between ethanol and ethylene, the key intermediate *$CH_2CHO$, which is bidentate bonded to the Cu surface through its C and O atoms, can either cleave a Cu−C bond to form $C_2H_4$ (the ethylene pathway) or break a C−O bond to yield $CH_3CHO$ (the ethanol pathway) (**Figure 6a**). DFT calculations indicate that Al doping enhances the overlap of O 2p and C 2p orbitals in *$CH_2CHO$ and increases the catalyst's oxophilicity, thus promoting ethanol production (**Figure 6b**). These findings are corroborated by in situ experiments and electrochemical tests, which report a FE of 84.5% for $C_{2+}$ products.[103]

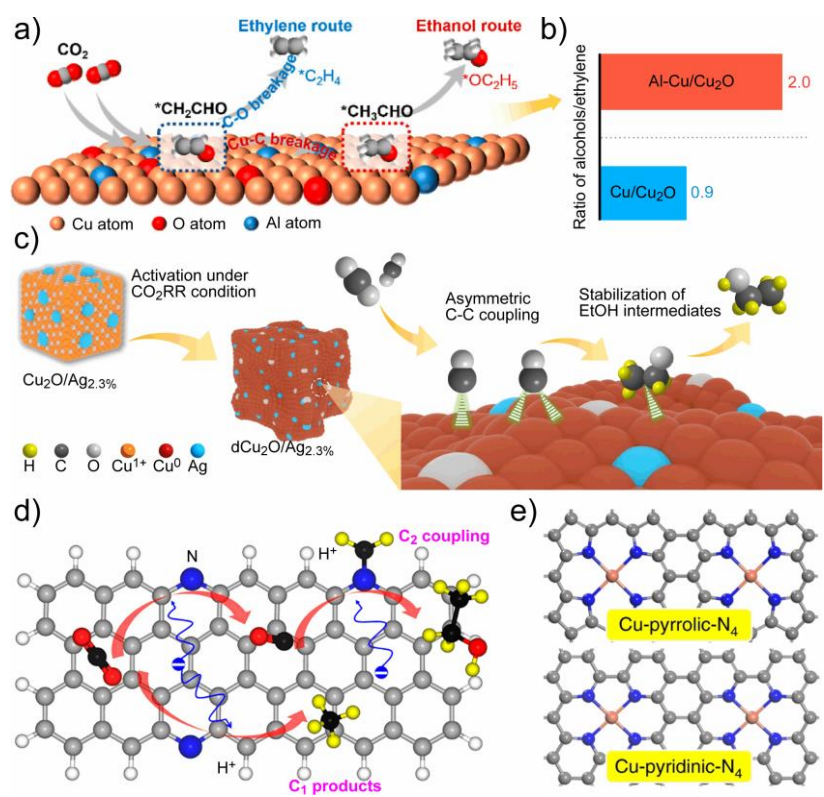

**Figure 6.** **a)** Diagram of the electrocatalytic CO$_2$RR to ethanol and ethylene. **b)** Effect of Al doping of Cu/Cu$_2$O on the selectivity of electrocatalytic CO$_2$RR. Figure a-b are reproduced with permission.[103] Copyright 2023, American Chemical Society. **c)** The ethanol production via asymmetric C−C coupling on Ag-doped Cu$_2$O catalysts. Reproduced with permission.[104] Copyright 2022, Springer Nature. **d)** The C−C coupling promoted by pyridine N edge. Reproduced with permission.[105] Copyright 2017, American Chemical Society. **e)** Schematics of the Cu-N$_4$ active site. Reproduced with permission.[110] Copyright 2020, Springer Nature.

With suitable metal dopants, the adsorption configurations of intermediates can be modified to stabilize key species, thereby enabling the production of specific products. For instance, Wang et al.[104] demonstrated that Ag doping into Cu$_2$O increases the proportion of *CO$_{bridge}$ adsorption configurations, allowing it to compete with the *CO$_{top}$ configuration. The *CO$_{bridge}$ configuration is particularly amenable to protonation, which promotes the formation of *CHO and subsequently triggers C−C coupling between *CO and *CHO after *CO$_{bridge}$ protonation (**Figure 6c**). DFT calculations indicate that Ag doping reduces the Cu coordination number, favoring ethanol production over ethylene owing to the more saturated and stable nature of the ethanol intermediate compared to that of C$_2$H$_4$. Experimental evidence supports this, as increased levels of *CO$_{bridge}$ and *OC$_2$H$_5$ (a crucial intermediate for ethanol formation) were observed upon Ag doping of Cu$_2$O.

Unlike metal doping, non-metal doping primarily modifies the electronic structure of the catalyst, as these heteroatoms typically do not directly participate in the electrocatalytic CO$_2$RR. However, in metal-free N-doped carbon materials, the N atom can serve as the main active site. For example, pyridinic N edges in graphene have been shown to effectively facilitate both the thermodynamic and kinetic conversion of CO$_2$ to *CO, making it a key site for C–C coupling (**Figure 6d**).[105] During the protonation process, H$^+$ primarily attacks the carbon atom in the intermediate, resulting in the formation of *CH$_2$. This is attributed to the low coordination of the C atoms bonded to the N site. Subsequently, *CH$_2$ couples with *CO on an adjacent carbon atom to produce the *CH$_2$CO intermediate, as shown in **Figure 3a**. This catalytic mechanism has been validated using N-doped graphene quantum dots (N-GQDs), which achieved a total Faradaic efficiency of 45% for ethylene and ethanol production.[106] In contrast, control experiments with pure GQDs and N-doped graphene oxide (GO, with fewer edge structures compared to GQDs) predominantly yielded hydrogen, carbon monoxide, and formate.[106]

In metal catalysts, non-metal doping is often used to regulate the coordination number of the active site. For instance, Wu et al.[107] employed this strategy by doping a copper-penetrating electrode with boron to create a B-Cu hybrid perovskite electrode (HPE), optimizing the Cu coordination number to 4.9. The B dopant distorts the Cu lattice, increasing the Cu–Cu bond length and resulting in unsaturated coordination. Additionally, the B atom withdraws electrons from adjacent Cu atoms, forming Cu$^{\delta+}$ (where 0 < δ < 1), which optimizes the adsorption strength and configuration of *CO, thus promoting C–C coupling. DFT calculations and in situ experiments indicate that the *CO$_{top}$ configuration becomes dominant with increased CO adsorption. This *CO$_{top}$ can readily be hydrogenated to *CHO, facilitating the asymmetric coupling of *CO$_{top}$ and *CHO on the B-Cu HPE. As a result, C$_{2+}$ product formation increases from 54.5% for the pure Cu-HPE to 78.9% with B doping.

Above C–C couplings are primarily facilitated by catalysts with densely packed active sites. In SACs, however, C–C coupling is inherently difficult due to the isolation of metal atoms, which limits the formation of C–C bonds and typically results in CO or HCOOH as the main products.[112-116] Recent studies, however, have shown that certain SACs can promote C–C coupling through appropriate doping treatments.[117-119] In this context, chemical doping plays a more significant role in modifying the chemical affinity and adsorption configurations compared to nanocluster catalysts. A notable example is the Cu–N$_4$ active center in the Cu–N–C catalyst (**Figure 6e**), which demonstrated an electrocatalytic CO$_2$RR leading to acetone with a maximum FE of 36.7%.[110] DFT simulations indicate that the free energy pathway from CO$_2$ to acetone is predominantly downhill, facilitating the reaction.[110] The C–C coupling in this Cu–N$_4$ system is primarily driven by the synergistic interaction between the Cu site and the coordinating pyrrole N, coupled with the restructuring of Cu–N bonds. Additionally, SACs can achieve C–C coupling through the reconfiguration of active sites into nanoclusters during the electrocatalytic CO$_2$RR. For instance, Xu et al.[120] reported approximately 91% selectivity for ethanol at -0.7 V$_{RHE}$ using Cu SACs supported on a carbon carrier. The active site in this case is a four-coordinate Cu$^{2+}$ that transforms into Cu$_3$ clusters during the reaction. The reaction mechanism follows the pathway CO$_2$ + CO$_2$ → HCOO* + HCOO* → CH$_3$ + H$_2$CO, where coupling between CH$_3$ and H$_2$CO results in ethanol formation, after which the active site returns to its original state. However, the process by which Cu clusters form in nearly pure SACs, without additional Cu species, remains unclear. The migration of single Cu atoms to form Cu clusters is inherently challenging due to the large distances between Cu atoms in most SACs. Overcoming this obstacle may require dense SACs, which present fabrication challenges but could potentially be realized on the surfaces of metastable materials with rich dangling bond states.[38]

The examples above illustrate that chemical doping is a potent strategy for controlling C–C coupling in catalysts. By introducing heteroatoms, the catalyst surface structure can be significantly modified, including changes to the metal oxidation state, the creation of new or altered active sites, modifications to the coordination environment, and enhancements to electrical conductivity. These changes, in turn, influence the adsorption behavior of key intermediates such as *CO, affecting their coverage and the subsequent hydrogenation processes. Some dopants, for instance, can stabilize the oxidation state of copper, especially $Cu^+$ species, lowering the protonation barrier and reducing overpotentials. This enhances the performance of Cu-based catalysts for electrocatalytic $CO_2RR$. However, designing chemical doping strategies that effectively promote $C_{2+}$ product formation requires careful integration of theoretical calculations and a thorough understanding of the $CO_2RR$ pathways, as illustrated in **Figures 2-4**.

### 4.3. Facet and phase engineering

Facet control refers to the strategic modification of the exposed facets of catalysts, a method that has proven highly effective in tuning catalytic performance for a variety of chemical reactions.[121-124] Different facets display distinct atomic arrangements and surface energies, which are directly related to the density of dangling bond states.[125-127] These differences significantly influence the activity and selectivity of the electrocatalytic $CO_2RR$. As a result, facet engineering remains one of the primary techniques for optimizing the electrocatalytic performance of Cu-based catalysts in electrocatalytic $CO_2RR$.

Low-index crystal facets, such as (100), (110), and (111), are characterized by relatively low Miller indices and feature regular atomic arrangements with smooth, flat surfaces and minimal defects.[128] These facets typically have low surface energies, which contributes to their stability during chemical reactions.[129] Early studies demonstrated that the Cu (100) facet is more prone to generating $C_2H_4$, whereas the Cu (111) facet shows a higher selectivity for $CH_4$.[130,131] The combination of the Cu (100) basal plane with Cu (111) steps also significantly promotes $C_2H_4$ production while suppressing $CH_4$ formation.[130,131] This led to the general understanding that Cu (100) is more favorable for the production of $C_2$ products (e.g., $C_2H_4$), while other low-index facets tend to favor $C_1$ products, such as HCOOH and $CH_4$.[132] The higher surface energy of Cu (100), coupled with its lower coordination number and increased dangling bond states, is thought to result in a higher binding affinity for intermediates.[38,80] This contributes to a lower energy barrier for *CO dimerization, which decreases further as the *CO coverage increases (as seen in the improved C−C coupling in **Figure 4a**). This observation was confirmed by Wang et al., who found that the Cu (100) facet is stabilized by $CO_2RR$ intermediates (**Figure 7a**).[133] Compared to the Cu (111) and Cu (211) facets, the Cu (100) facet is more favorable for electrocatalytic $CO_2RR$. DFT calculations indicate that the adsorption of $CO_2RR$ intermediates is significantly stronger on the Cu (100) facet, as compared to HER-related intermediates, and stronger than on the Cu (111) and Cu (211) facets.[133] This stronger adsorption leads to the lowest *CO−*CO coupling barrier on the Cu (100) facet. Similarly, Zhang et al. reported similar results in comparing electrocatalytic $CO_2RR$ on Cu (100) and Cu (111) facets.[134] Notably, *CO adsorbs predominantly on the Cu (100) facet in a top adsorption configuration, while it adopts a bridge adsorption configuration on the Cu (111) facet (**Figure 7b**). For the $CO_2 \rightarrow C_2H_4$ reaction, the rate-determining step on Cu (100) is C−C coupling, whereas on Cu (111), the protonation of *CO with $H_2O$ is the rate-determining step, leading to distinct catalytic properties for each facet (**Figure 7b**).

In contrast to low-index crystal facets, high-index facets possess atoms with lower coordination numbers, leading to a more widespread distribution of dangling bond states and consequently higher binding affinities.[135] This characteristic makes high-index facets more favorable for C−C coupling, as shown in **Figure 4a**. However, the higher surface energy of these facets presents challenges in their fabrication and stabilization, making them inherently more difficult to work with than low-index facets. Kim et al. addressed this issue by creating a wrinkled Cu surface that includes high-index facets such as (200), (210), and (310), with the (310) facet being the most prominent.[136] This surface was synthesized using chemical vapor deposition (CVD) of graphene, followed by detachment, as illustrated in **Figures 7c**. These high-index facets are stabilized by a high density of step sites on the Cu substrate (**Figure 7d**). The exposure of these facets enhances the surface binding affinity, resulting in significantly improved $C_2$ production, particularly for ethanol ($C_2H_5OH$). DFT calculations show that the Cu (310) facet exhibits a lower activation energy barrier for C−C coupling compared to (200) and (210) facets, and it is thermodynamically more favorable for the formation of $C_2H_5OH$ rather than $C_2H_4$.

By forming interfaces between different facets, electronic polarization and charge redistribution can be induced, similar to the behavior observed at the crystal boundaries between mismatched regions in 2D materials, which enhances chemical affinity.[82,83,137] For example, a catalyst with rich interfaces between CuO and $Al_2CuO_4$ has been shown to achieve a remarkable 82.4% production efficiency for $C_2H_4$ (**Figure 7e**).[138] In this structure, charge transfer from Cu to Al occurs on $Al_2CuO_4$, leading to an increase in the average oxidation state of Cu.[138] DFT calculations suggest that *CO is preferentially adsorbed on the $Al_2CuO_4$ facet, while CO is highly prone to be produced on CuO and transferred to $Al_2CuO_4$. Notably, a single Cu atom on $Al_2CuO_4$ can bind two *CO molecules. This is attributed to the tetrahedral geometry of Cu in the catalyst, which results in smaller d-orbital splitting (stronger binding to CO) compared to the square planar geometry. Additionally, the charge difference map reveals an accumulation of charge between the two C atoms (**Figure 7f**), which aids in the C−C coupling during the subsequent reaction, thereby promoting the production of $C_{2+}$ products. This structure can also be considered as a hybrid structure, which will be discussed further in **Section 4.5**.

An extreme example of facet engineering is found in high-entropy alloys (HEAs), where the size mismatch between different elements and the resulting lattice distortion provide numerous opportunities to diversify atomic coordination structures. These structural variations are key in lowering overpotentials and enhancing the selectivity of reactions by breaking the intrinsic adsorption energy scaling relations in complex reaction pathways.[139,140] However, optimizing these HEAs often requires substantial computational resources, including DFT simulations and machine learning (ML). For instance, Chen et al.

designed an HEA of $Fe_{0.2}Co_{0.2}Ni_{0.2}Cu_{0.2}Mo_{0.2}$ with the help of DFT simulations and ML.[141] They predicted the adsorption energies of 1280 possible adsorption sites for intermediates (COOH*, CO*, and CHO*) involved in $CO_2$ and CO protonation steps (the bottom pathway in **Figure 2c**). By rotating COOH* and CHO* on the catalyst surface, they circumvented the scaling relation between the adsorption energies of these intermediates (**Figure 7g,h**), leading to highly efficient $CO_2RR$ with a limiting potential of 0.29-0.51 $V_{RHE}$—significantly lower than the theoretical value of 0.7 $V_{RHE}$ on the volcanic plot.[48,142,143] This improvement is further validated by Nellaiappan et al., who demonstrated AuAgPtPdCu HEA nanocrystals capable of producing 100% gaseous products (CO, $CH_4$, $C_2H_4$, and $H_2$) in electrocatalytic $CO_2RR$, with a significantly reduced limiting potential compared to pure copper.[144] Unfortunately, most studies in this field have focused on $C_1$ products, and there are few cases demonstrating the efficient production of $C_{2+}$ products with HEAs.

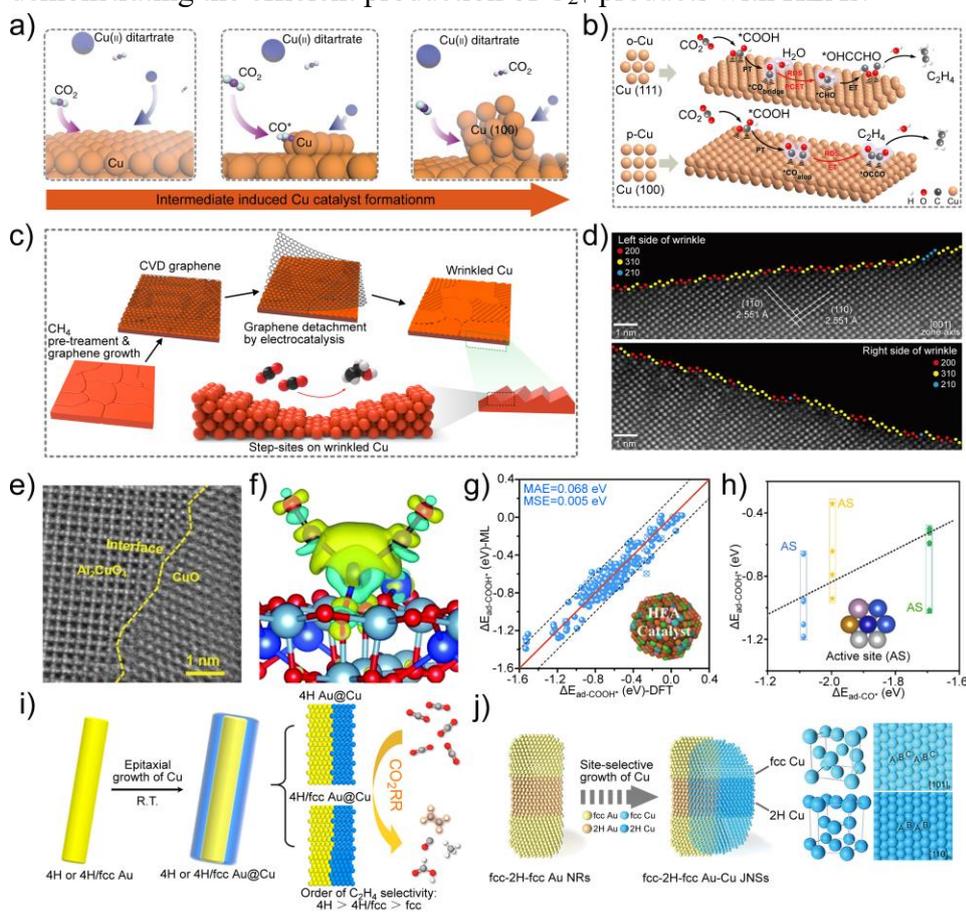

**Figure 7**. **a)** Regulation of Cu surface with $CO_2RR$ intermediates. More Cu (100) facet is formed with the $CO_2RR$ intermediates as capping agents. Reproduced with permission.[133] Copyright 2019, Springer Nature. **b)** $CO_2RR$ on Cu (111) and Cu (100) facets. Reproduced with permission.[134] Copyright 2024, National Academy of Sciences, USA. **c)** Fabrication of Cu substrate with high-index facets. **d)** Transmission electron microscopy (TEM) images of the high-index facets. Figures c-d are reproduced with permission.[136] Copyright 2021, American Chemical Society. **e)** TEM image of $CuO/Al_2CuO_4$. **f)** The charge distribution of two CO bind with Cu in $Al_2CuO_4$ (yellow: charge accumulation; cyan: charge depletion). Al, Cu, O, C are indicated as sky blue, blue, red, and brown color, respectively. Figures e-f are reproduced with permission.[138] Copyright 2022, Royal Society of Chemistry. **g)** Plots of the adsorption energy of COOH* predicted with ML against that of DFT calculated. The accuracy was evaluated using mean absolute error (MAE) and mean square error (MSE) between above two data categories. **h)** Correlation between the adsorption energy of COOH* and CO* on active site. Figures g-h are reproduced with permission.[141] Copyright 2022, American Chemical Society. **i)** The growth of 4H or 4H/fcc Cu. Reproduced with permission.[153] Copyright 2020, American Chemical Society. **j)** The preparation of fcc-2H-fcc Au-Cu. Right side: the unit cell and basal plane of fcc/2H Cu. Reproduced with permission.[154] Copyright 2024, Wiley.

Similar to facet engineering, phase engineering focuses on exposing facets with regulated electronic structures to enhance catalytic performance.[145-149] Phase engineering has been extensively applied to 2D materials for electrochemical energy storage and HER, as metastable phases (e.g., 1T $MoS_2$) offer a significantly enhanced electronic state density near the Fermi level.[82,83] The chemical affinity of these phase-engineered 2D materials to *CO remains insufficient for efficient $CO_2$RR toward $C_{2+}$ products.[150-152] Nevertheless, the production of $C_{2+}$ products with phase engineered Cu-based catalysts is possible. For example, **Figure 7i** compares face-centered cubic (fcc) Cu with the 4H Cu phase. DFT calculations suggest that the 4H phase and the 4H/fcc Cu interface have lower energy barriers for $CO_2$ reduction to *CO compared to pure fcc Cu.[153] These phase-engineered surfaces or interfaces more readily form *CHO intermediates, facilitating the *CO−*CHO asymmetric coupling to yield $C_2H_4$ products.[153] In a related study, Ma et al. fabricated fcc-2H-fcc Au-Cu Janus nanostructures (JNSs) through epitaxial growth using crystal seeds of fcc-2H-fcc Au nanorods (NRs), exposing fcc-2H-fcc Cu on the surface (**Figure 7j**).[154] The resulting catalyst exhibited FEs of 55.5% and 84.3% for ethylene and multicarbon products, respectively. The presence of both 2H-Au and 2H-Cu phases promoted a diverse range of *CO adsorption configurations, including both *$CO_{top}$ and *$CO_{bridge}$, which facilitated the C−C coupling. Additionally, the d orbitals of Cu and Au in the 2H phase are positioned closer to the Fermi level, enhancing the chemical affinity of the catalyst. On the 2H Cu/Au surface, the s- and p-orbitals of $CO_2$RR intermediates—from $CO_2$ reduction to C−C coupling and subsequent hydrogenation—showed a good linear correlation, gradually down-shifting from the Fermi level and lowering the energy barriers for the reduction process, particularly for C−C coupling.[154] This Janus structure and its electronic polarization effects, induced by asymmetric facets, will be further discussed in **Section 4.5**.

The examples discussed above demonstrate that selectivity of electrocatalytic $CO_2$RR can be effectively regulated toward $C_{2+}$ products through facet and phase engineering. Phase engineering typically involves fabricating unconventional Cu phases and/or combining them with other metallic phases to modulate the electronic structure and catalytic properties. In facet engineering, C−C coupling is often achieved by utilizing high-index facets, which provide a stronger binding to *CO compared to low-index facets (see **Figure 4a**). This enhanced coupling can also occur at the interface of two different facets, where charge transfer between phases further improves the binding affinity for intermediates. While phase engineering has led to some success in promoting $C_{2+}$ production, relatively few cases have been reported in the context of Cu-based catalysts. One of the challenges in facet engineering is the difficulty in developing reliable protocols to access and stabilize high-index facets. Furthermore, phase engineering, especially in the case of HEAs, demands significant computational resources, including extensive DFT calculations and ML models, due to the large number of metal interlayers and possible phase combinations involved in optimizing C−C coupling.

### 4.4. Strain and relaxed bonding structures
Strain and relaxed bonding structures, often resulting from lattice mismatches, have been extensively studied in the field of catalysis.[79,84,155-157] In metal surfaces, these strain-induced effects can significantly influence the position of the d-band center, which in turn alters the binding affinity of adsorbates, thus enhancing catalytic performance.[158] For late transition metals, such as Cu, Ni, and Pt (which are commonly used in catalytic applications), previous studies have identified general trends: tensile lattice strain causes an upshift in the d-band center, leading to stronger interactions with adsorbates (as illustrated in **Figure 1b**), whereas compressive strain results in a downshift, weakening the interaction by moving the d-band away from the Fermi level.[159] Relaxed bonding structures, similar to tensile strain, enlarge the atomic spacing, exposing more dangling bond states that can enhance chemical bonding and, consequently, improve catalytic activity.

Strain and relaxed bonding structures have the potential to disrupt the scaling relationships involved in the electrocatalytic $CO_2RR$. As discussed in **Section 4.3**, HEA materials have been shown to break the scaling relationships between the adsorption energies of COOH*, CO*, and CHO* intermediates. Similarly, uniaxial tensile strain applied to Cu (100) and Cu (211) facets has been found to increase the adsorption energies of *COOH and *CHO intermediates, while decreasing the adsorption energies of CO* and *COH intermediates.[160] This contrasting response in the adsorption energies of different $CO_2RR$ intermediates indicates a breaking of the traditional scaling relationships in electrocatalysis. The underlying cause of this behavior has been attributed to the differential strains induced by the various adsorbates and their respective adsorption configurations.[160]

Chang et al. fabricated various Ag-Cu core-shell nanoparticles (NPs) by gradually depositing a Cu shell onto an Ag core, which significantly improved the selectivity for hydrocarbons in electrocatalytic $CO_2RR$ (**Figure 8a**).[161] Upon forming the Cu cladding, Cu dominated as the electrocatalytic reaction center. The lattice mismatch between Ag and Cu increased the interatomic distance of Cu atoms compared to bulk Cu, inducing tensile surface strain that altered the adsorption of $CO_2RR$ intermediates.[161] The larger atomic radius of Ag, relative to Cu, caused a stretching of the Cu lattice, which upshifted the Cu d-band center toward the Fermi level, enhancing its binding affinity for *CO and facilitating the formation of hydrocarbons.[161] This tensile strain increased $C_2H_4$ selectivity from approximately 0% to 28.6%.[161] Li et al. also reported a tensile-strain Cu structure, where Cu nanosheets were coated on a Cu high-entropy perovskite (NSL-Cu HPE) using a chemical etching and electroreduction technique.[162] The resulting Cu lattice distortion in the NSL-Cu HPE induced relaxed Cu-Cu bonding and unsaturated Cu coordination (**Figure 8b**).[162] DFT calculations confirmed that the tensile strain upshifted the Cu d-band center closer to the Fermi level, increasing the binding strength of *CO and slightly enhancing the binding strength of *H. This strain not only promoted *CO adsorption but also facilitated the multi-step electron-proton coupling process, especially the hydrogenation of *CO in $CO_2RR$.[162] The improved *CO coverage and the lowered energy barrier for the asymmetric coupling of *CO−*CHO (or COH) led to a high selectivity for $C_{2+}$ products (**Figure 8b**).[162] Notably, the enhanced asymmetric C−C coupling created an unbalanced coordination environment that favored the formation and stabilization of $CH_3CHO$ (**Figure 3b**), a key intermediate for $C_2H_5OH$ and n-propanol formation. This resulted in the production of n-propanol through a CO insertion reaction (**Figure 3e**).

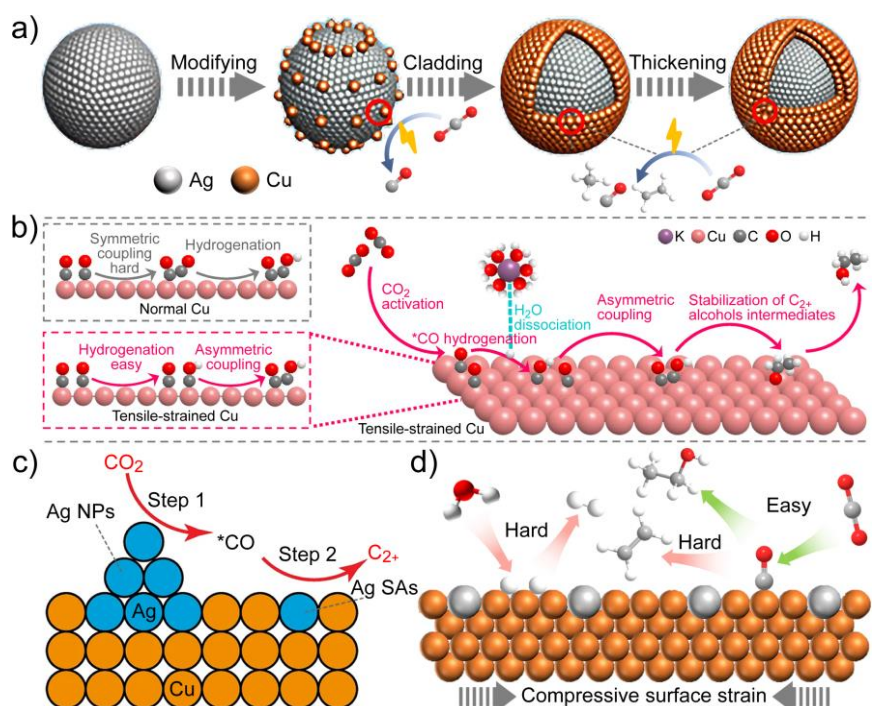

**Figure 8. a)** Fabrication of Ag-Cu core shell structure, to regulate the CO$_2$RR selectivity. Reproduced with permission.[161] Copyright 2017, American Chemical Society. **b)** Schematics of the boosted C$_{2+}$ and C$_{2+}$ alcohols generation over NSL-Cu HPE. Reproduced with permission.[162] Copyright 2024, Wiley. **c)** The cascade catalysis mechanism over AgCu SANP. Reproduced with permission.[163] Copyright 2023, Springer Nature. **d)** Schematics of strain compression applied on CuAg alloy to tune the catalytic selectivity. Reproduced with permission.[164] Copyright 2017, American Chemical Society.

In addition to tensile strain, compressive strain also significantly influences the electrocatalytic CO$_2$RR performance of catalysts. For example, AgCu SANP catalysts, which combine Ag NPs and Ag SACs on Cu NPs, can achieve an impressive 94% FE for C$_{2+}$ products.[163] In this system, CO$_2$ is first reduced to CO on the Ag NPs, and asymmetric C−C coupling occurs in the high concentrations of CO, facilitated by the synergistic interaction between Ag SACs and surrounding Cu atoms. The larger atomic radius of Ag compared to Cu induces compressive strain on the Cu lattice, shortening Cu−Cu bonds (**Figure 8c**). The asymmetric Cu sites adjacent to the Ag sites serve as the actual active sites for C−C coupling. This AgCu SANP surface exhibits lower energy barriers for *CO hydrogenation (*CO → *CHO) and *CO−*CHO coupling than the Cu (100) surface, contributing to the efficient production of C$_{2+}$ products.[163] Although compressive strain typically weakens the binding affinity of intermediates, as shown in **Figure 4a**, appropriate regulation of competitive adsorption can alter this outcome. For instance, CuAg alloys under compressive strain shift the valence band structure of Cu toward higher energy levels, weakening the binding of H and O relative to CO, and promoting the selective generation of CO-derived products.[164] Additionally, the decreased adsorption of H and the reduced oxophilicity of compressively strained Cu enhance the selectivity for carbonyl-containing C$_{2+}$ products (**Figure 8d**).[164] Using this CuAg alloy, Clark et al. demonstrated that HER was suppressed, and the FE for C$_{2+}$ products increased from 56% to 94%, compared to pure Cu, during electrocatalytic CO$_2$RR.[164]

The studies above suggest that the electronic structure — and, by extension, the catalytic properties — of a material can be effectively tuned by managing the surface strain in the catalyst. These strain-induced changes in catalytic activity can be largely explained by general trends in chemical affinity and synergistic effects within complex metal systems. It is important to note, however, that these trends in chemical affinity apply primarily to late transition metals, where the d-band is more than half-filled. Early transition metals, with less than half-filled d-bands, exhibit opposite behavior — showing lower adsorption energies upon lattice expansion, in contrast to the typical trend observed in late transition metals.[165] This difference can be understood through the classical d-band model. Specifically, when the surface strain is applied, the expansion reduces the overlap of the wavefunctions, which narrows the metal d-band. In late transition metals (e.g., Cu, Ni, Pt), this band narrowing leads to an increased population of the d-band, upshifting the d-band center to maintain the degree of d-band filling. In early transition metals (e.g., Sc, Ti, V), however, the same mechanism downshifts the d-band center. A deeper understanding of the principles governing strain and relaxed bonding structures on the electronic properties of catalysts is crucial for designing catalysts with enhanced and targeted performance.

### 4.5. Hybrid structures and surface functionalization

The formation of $C_{2+}$ products from electrocatalytic $CO_2$RR can be promoted by controlling internal and surface composition of the catalyst, which is attributed to various effects such as synergistic,[166] alloying,[167] tandem,[168] chemical state,[169] and microstructural effects.[170] In Janus-like heterostructures, the modulated electrocatalytic $CO_2$RR is usually occurs at the interface, owing to the local charge polarization and distorted structures at the interface. This situation may be changed if the hybrid structure is thin enough, giving the surface electronic structure affected by the charge polarization in the interlayer. Nevertheless, the regulation of electrocatalytic $CO_2$RR on the ultrathin 2D hybrid structure is rarely discussed.

**Figure 9a** illustrates an example of the Au-Cu heterostructures (particle size: 80-150 nm, Cu thickness: < 20 nm), which is formed from a dimer with controlled Au exposure, namely the Au-Cu$_I$, Au-Cu$_{II}$ and Au-Cu$_{III}$.[171] These controlled heterostructures have the main electrocatalytic $CO_2$RR products of methanol, ethanol, and ethylene, respectively.[171] The contacted interface between Au and Cu increases from Au-Cu$_I$ to Au-Cu$_{II}$, then decreased in the Au-Cu$_{III}$. These interactions affected the valance of Cu, giving the Cu−Cu coordination numbers of 8.1, 7.7, and 6.6 for Au-Cu$_I$, Au-Cu$_{II}$, and Au-Cu$_{III}$, respectively.[171] Besides, the $HCO_3^-/CO_3^{2-}$ ratios at the surface of Au-Cu$_{II}$ and Au-Cu$_{III}$ decreased sharply with increasing current density.[171] These two factors resulted in a localized high pH on the surface, favoring for C−C coupling to produce $C_{2+}$ products of the Au-Cu$_{II}$ and Au-Cu$_{III}$.[172] Compared to the Au-Cu$_{III}$ with predominate *CO$_{bridge}$ adsorption configuration, the Au-Cu$_{II}$ has higher *CO coverage to give the *CO$_{top}$ adsorption configuration on the surface (**Figure 9a**). This explained the transformation of the ethylene to ethanol (also see **Figure 6c** for the same transformation with different *CO adsorption configurations), catalyzed by Au-Cu$_{III}$ and Au-Cu$_{II}$, respectively.

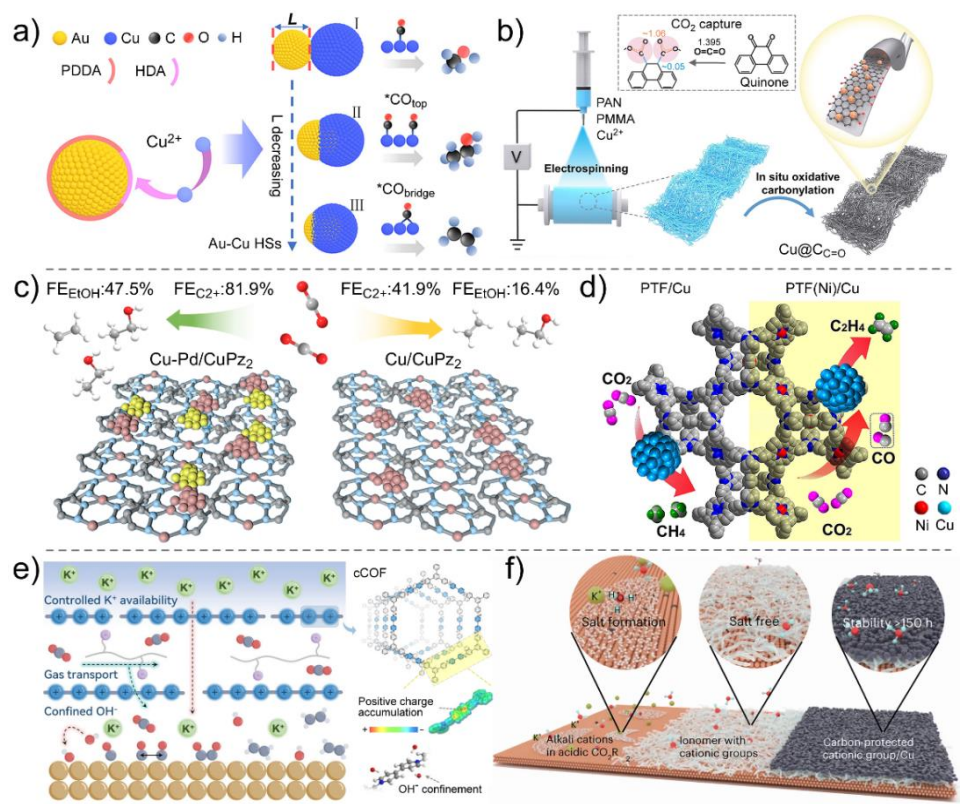

**Figure 9. a)** Au-Cu heterostructures with regulated $CO_2RR$ selectivity. HAD: 1-hexadecylamine; PDDA: poly-(diallyldimethylammonium) chloride. Reproduced with permission.[171] Copyright 2024, Elsevier. **b)** The preparation of $Cu@C_{C=O}$ with enhanced $CO_2$ capture and *CO coverage. Inset shows LBO of $CO_2$ capture with and without quinone functionalization. PAN: polyacrylonitrile; PMMA: polymethylmethacrylate. Reproduced with permission.[178] Copyright 2024, Wiley. **c)** The comparison of $Cu/CuPz_2$ and $Cu-Pd/CuPz_2$ in electrocatalytic $CO_2RR$. Reproduced with permission.[173] Copyright 2024, Wiley. **d)** The comparison of PTF/Cu and PTF(Ni)/Cu in electrocatalytic $CO_2RR$. Reproduced with permission.[53] Copyright 2021, Wiley. **e)** The cCOF/PFSA interaction at the surface, structure of cCOF, electrostatic potential map of the viologen unit, and simulation depicting the $OH^-$ confinement around the cCOF. Reproduced with permission.[181] Copyright 2024, American Chemical Society. **f)** The surface structures of blank Cu electrode, CG coated Cu electrode, and Cu electrode with a carbon layer. Reproduced with permission.[185] Copyright 2023, Springer Nature.

Xie et al. have demonstrated the promotion on the production of $C_{2+}$ products of the electrocatalytic $CO_2RR$, by complexing the oxyphilic metal Pd with Cu.[173] As illustrated in **Figure 9c**, the Cu-Pd bimetallic cluster is bridged on $CuPz_2$ (Pz = pyrazole).[173] The inclusion of Pd species enhances interfacial electron transfer, facilitates *CO adsorption and *CO−*CO dimerization.[173] Additionally, the oxophilicity of Pd and the synergistic tandem effect of Cu-Pd dual sites stabilizes the adsorption of $CH_2CHO^*$, thereby preventing the formation of ethylene and promoting the conversion to ethanol (the route follows the orange arrows in **Figure 3b**).[173] As a result, the FE of the $C_{2+}$ products was improved to 81.9% (FE of EtOH: 47.5%) from 41.9% (FE of EtOH: 16.4%) of the catalyst with Cu cluster loaded alone (**Figure 9c**).[173] Similar promotions on the production of EtOH have been demonstrated on Cu-Pd core-shell nanocatalyst and Cu-Pd subnanoalloys.[174,175]

Aside from above materials, MOFs and covalent organic frameworks (COFs) have also been applied to build hybrid structure to promote the production of $C_{2+}$ products.[53,176] For example, Cao et al. prepared a catalyst with atomically isolated Ni-N sites anchored on porphyrin triazine framework [PTF(Ni)], which efficiently catalyzes the conversion of $CO_2$ to CO with a FE of 94-98%.[177] By depositing Cu NPs on the PTF(Ni) to form PTF(Ni)/Cu, a FE of 57.3% for the production of $C_2H_4$ was demonstrated.[53] This value is 6-fold higher than the maximal FE (9.6%) of the control catalyst without the presence of Ni (PTF/Cu, **Figure 9d**).[53] In PTF/Cu, the rate-determining C−C coupling step for the formation of $C_{2+}$ products on Cu NPs has a free energy of 0.97 eV versus 0.94 eV of the rate-determining step for the formation of $CH_4$ (*CO→*CHO), giving the predominant formation of $CH_4$.[53] By contrast, in the PTF(Ni)/Cu, a high concentration of CO is generated on PTF(Ni), which is rapidly transferred to nearby Cu NPs (**Figure 9d**), leading to a high *CO coverage to reduce energy barrier of *CO−*CO coupling to give high FE of $C_2H_4$ (also see **Figure 4a**).[53] Unlike traditional metal compound, the MOF/COF materials have significant large amount of varieties with porous structure, giving broad potentials in building hybrid structures for various $C_{2+}$ products.

While above hybrid structures are rely on the electronic polarization between interlayers, the modification of catalyst can directly regulate the surface nature to alter the electrocatalytic $CO_2RR$. **Figure 9b** illustrates an example where the Cu sites are immobilized on the quinone functionalized carbon nanofibers ($Cu@C_{C=O}$).[178] Although the functionalization did not happened on the Cu sites, the electrocatalytic $CO_2RR$ can still be regulated by facilitating the $CO_2$ capture and activation owing to two carbonyl groups in the quinone.[179] Upon bonding with quinone, the Laplacian bond order (LBO) of the C−O bond strength in $CO_2$ was reduced by approximately 24% (**Figure 9b**).[179] The weak interaction between quinone and $CO_2$ facilitates the dissociation of the activated *$CO_2$ intermediate (**Figure 9b**), i.e., *CO, which readily migrates to the nearby active Cu sites for further catalysis. A mechanism similar to the PTF(Ni)/Cu (discussed above in **Figure 9d**) subsequently occurs, giving high *CO coverage to promote the formation of $C_2H_4$.

In addition to increasing the *CO coverage to allow more efficient catalysis, the $C_{2+}$ yield can also be improved by enhancing the catalyst binding with *CO, upon suitable functionalizations. For example, Ag-Cu catalysts dip coated with thiadiazole ($N_2SN$) and triazole ($N_3N$) were able to enhance the FE of $C_{2+}$ products of electrocatalytic $CO_2RR$ from approximately 20% to 80%.[180] The $N_2SN$ and $N_3N$ groups have electron-withdrawing properties, leading to the formation of $Cu^{\delta+}$ ($0 < \delta < 1$).[180] $Cu^{\delta+}$ facilitates the formation of *$CO_{top}$, a known key intermediate involved in the C−C coupling step.

Actually, the surface functionalization can create microenvironment at the electrode-electrolyte interface to improve the electrocatalytic $CO_2RR$. It is known that the ion migration at the cathode under the electric field can lead to unstable alkalinity and carbonate precipitation, degenerating the selectivity and activity of the electrocatalytic $CO_2RR$. To address these problems, Fang et al. proposed a "charge release" strategy by incorporating tiny amounts of oppositely charged anionic ionomers (e.g., perfluorinated sulfonic acid, PFSA) into the cationic COF (cCOF/PFSA) on the Cu surface (**Figure 9e**).[181] This functionalization effectively releases the hidden positive charge within the cCOF, enhancing surface immobilization of cations to impede both outward migration of $OH^-$ and inward migration of cations.[181] As a result, the carbonate precipitation is inhibited, and strong alkaline microenvironment is created (**Figure 9e**).[181] Notably, the ionomer's hydrophobic chains facilitate efficient $CO_2$ transport. This hydrophobic strong alkaline microenvironment optimizes the strength and configuration of adsorbed *CO, giving the C−C coupling to form the $C_{2+}$ product through the combination of $*CO_{top}$ and $*CO_{bridge}$.[181] Compared to bare Cu control electrode (Stark tuning slope: 9.3 $cm^{-1}$ $V^{-1}$), the cCOF/PFSA-modified catalyst exhibited a significant improved Stark tuning slope (27 $cm^{-1}$ $V^{-1}$), a phenomenon of stronger surface electric field strength.[181] This enhanced electric field is known to stabilize the $CO_2RR$ intermediates.[182-184] Fan et al. have also demonstrated an approach to stabilizing $CO_2RR$ intermediates with the enhanced electric field, by functionalizing a thin layer of ionomer with immobilized cationic groups (CG) at the Cu surface (**Figure 9f**).[185] This electric field lowers the proton diffusion rate and increases the local pH for selective $C_{2+}$ products. By further adding a carbon layer to prevent the breakdown of CG layer, a high FE of 80% for the $C_{2+}$ production was achieved in the strong acid environment for a 150 h operation (**Figure 9f**).[185]

The above cases show that constructing hybrid structures and functionalized surfaces are widespread strategies to facilitate the conversion of $CO_2$ to $C_{2+}$ products. These strategies typically alter the electronic structure at the surface or interface, tuning the adsorption behavior of key intermediates for selective $C_{2+}$ production. Functionalized surfaces can also improve the local microenvironment to achieve high selectivity by increasing local pH, inhibiting HER and stabilizing key intermediates. In addition, a stable local environment plays as an imperative role in stabilizing the catalysis.

## 5. Summary and Perspectives

### 5.1. Summary of the C–C coupling and engineering techniques

In this report, we have comprehensively discussed recent advances in catalyst design for electrocatalytic $CO_2RR$ targeting high-value $C_{2+}$ products, with a particular focus on the critical factors influencing C–C coupling and catalyst engineering strategies. The efficiency of C–C coupling and the catalytic selectivity for $C_{2+}$ products are intricately tied to the strength and configuration of *CO adsorption, as well as the catalyst's affinity for specific atoms (e.g., C and O). The following key insights have been derived:

- **Binding affinity and active sites:** The formation of $C_{2+}$ products, as opposed to $C_1$ products, requires a stronger binding affinity to stabilize *CO, coupled with an abundance of catalytic active sites to minimize the distance between intermediates and facilitate C–C coupling. However, overly strong *CO binding can result in catalyst poisoning, hindering the reaction.
- **Role of CO* as a key intermediate:** The generation of *CO through $CO_2$ activation and adsorption via the carbon atom is a prerequisite for C–C coupling.
- **Adsorption Configurations:** The *CO molecule can adopt multiple adsorption configurations, including $*CO_{top}$, $*CO_{bridge}$, and $*CO_{hollow}$. Optimizing *CO coverage lowers the energy barrier for C–C coupling by shifting the adsorption configuration from $*CO_{bridge}$ to $*CO_{top}$. A moderate *CO coverage not only reduces the coupling energy barrier but also suppresses the HER, thereby enhancing the efficiency of $C_{2+}$ product formation.
- **Hydrogenation of CO***: Multi-step hydrogenation of *CO favors the formation of $C_1$ products, as the hydrogenation of $C_1$ intermediates is kinetically more favorable than C–C coupling. To promote C–C coupling, it is crucial to minimize excessive hydrogen adsorption on the catalyst surface.
- **Competitive adsorption of intermediates:** During later stages of electrocatalytic $CO_2RR$, competitive adsorption of intermediates via either C or O atoms influences product distribution. For instance, starting from the CHCHO* intermediate, ethylene or ethanol is likely to form depending on whether the catalyst stabilizes the C atom (facilitating proton attack on the O atom) or the O atom (facilitating proton attack on the C atom). Enhanced binding strength to either atom may also promote C–O bond dissociation, favoring ethylene production.
- **Asymmetric coupling:** *CO hydrogenation intermediates, such as *CHO, preferentially undergo asymmetric coupling with *CO. This mechanism reduces the energy barrier compared to symmetric *CO–*CO coupling.
- **Stabilization of $C_{2+}$ intermediates:** Stabilization of $C_{2+}$ intermediates can be achieved through electric field effects and local pH adjustments, which support C–C coupling and the subsequent formation of $C_{2+}$ products.

These insights underscore the nuanced interplay of *CO adsorption strength, intermediate configurations, and catalyst design in achieving selective and efficient electrocatalytic $CO_2RR$ to $C_{2+}$ products. Future research should focus on fine-tuning these parameters, leveraging advanced computational and experimental tools to unlock new pathways for catalyst optimization and high-value product synthesis.

To enable efficient C−C coupling during the electrocatalytic $CO_2RR$ for $C_{2+}$ product formation, various catalyst engineering strategies have been explored, including chemical doping and SACs, phase engineering, facet control, defect introduction, lattice strain modulation, relaxed bonding, hybrid structures, and surface functionalization. The primary principles behind these strategies are summarized as follows:
- **Defected structures:** The high density of unsaturated sites in defected structures makes them highly effective in facilitating C−C coupling. However, these structures are often thermodynamically unstable, and a high defect density can compromise electrical conductivity and alter adsorption configurations, potentially affecting catalytic selectivity. For example, excessive binding of the $CH_2CHO*$ intermediate via the O atom may promote C–O bond dissociation, shifting the product from $CH_3CH_2OH$ to $C_2H_4$.
- **Chemical doping:** Active metal dopants can enhance *CO adsorption and coverage, increasing the likelihood of C−C coupling. Additionally, metal dopants with distinct chemical affinities compared to the pristine catalyst can enable asymmetric C−C coupling, lowering energy barriers and allowing for selective $C_{2+}$ product synthesis. Non-metal dopants primarily modulate the coordination number of active sites, optimizing *CO adsorption strength, coverage, and configuration to promote C−C coupling.
- **High-index facets:** Catalysts with high-index facets possess a high density of atomic steps, kinks, and low-coordination atoms, which exhibit strong chemical affinity for

*CO, enhancing C−C coupling. However, stabilizing and fabricating such high-index facets remains a challenge. For phase-engineered Cu-based catalysts, enhanced binding strength to *CO and increased dangling bond states on the surface facilitate C−C coupling, particularly at phase interfaces that promote asymmetric coupling.

- **Strain engineering:** For late transition metals, tensile lattice strain enhances C−C coupling, while compressive strain suppresses it. In early transition metals, the trend is reversed, with tensile strain suppressing and compressive strain promoting C−C coupling, attributed to differences in d-band filling.
- **Hybrid structures:** Charge polarization and structural distortion in hybrid catalysts directly strengthen *CO binding, improving C−C coupling. Synergistic, tandem, and microstructural effects contribute to enhanced electrocatalytic $CO_2RR$ performance. Surface functionalization further stabilizes intermediates for $C_{2+}$ products by creating tailored electrode-electrolyte microenvironments in addition to improving *CO affinity.
- **HEAs:** HEAs offer numerous active sites and modify adsorption configurations, breaking scaling relations and enhancing C−C coupling. Despite their promise, electrocatalytic $CO_2RR$ on HEAs for $C_{2+}$ products remains underexplored, with significant computational resources (e.g., DFT simulations, ML models) needed for rational HEA design.
- **SACs:** Traditionally, SACs have been considered unsuitable for C−C coupling due to the isolated nature of metal atoms. However, suitable chemical doping can alter adsorption configurations and improve chemical affinity, enabling efficient C−C coupling. Highly dense SACs may theoretically facilitate C−C coupling, but their fabrication presents considerable challenges, limiting current research in this area.

These engineering strategies demonstrate the potential of precise catalyst design to optimize C−C coupling and drive the selective production of $C_{2+}$ products. Future investigations should prioritize addressing the inherent limitations of these approaches, leveraging both experimental advancements and computational insights to realize their full potential.

## 5.2. Advanced characterization, theoretical insights, and practical challenges

Beyond improving *CO binding affinity, the discussed engineering strategies emphasize the regulation of adsorption configurations, asymmetric coupling, and microenvironment optimization to facilitate C–C coupling and reduce the energy barriers of electrocatalytic $CO_2RR$ toward $C_{2+}$ products. Directly predicting these factors in engineered catalysts remains a formidable challenge. However, with the integration of theoretical simulations, advanced characterization techniques, and in situ experiments, the complex relationships between catalyst properties and these critical factors can be elucidated to inform more precise catalyst design for $C_{2+}$ products.

For structural characterization, high-resolution TEM (HRTEM) techniques, such as aberration-corrected scanning TEM (AC-STEM), are indispensable. In most cases, general HRTEM or AC-STEM provides sufficient detail about the fine structure of catalysts. However, for catalysts with structures sensitive to electron beams, such as ultrathin 2D MOF materials, cryogenic TEM (cryo-TEM) with low acceleration voltage or double aberration-corrected TEM at minimal electron doses is required.

Additional structural details, such as bond lengths and coordination environments, can be obtained using synchrotron-based X-ray absorption fine structure (XAFS) techniques, which are crucial for understanding the chemical affinity of catalysts. Beyond structural insights, in situ characterization tools such as in situ infrared absorption spectroscopy and Raman spectroscopy are vital for monitoring electrocatalytic $CO_2RR$ intermediates and catalytic pathways. These techniques help unravel the effects of electrolyte components (e.g., anions and cations) on catalytic activity and the evolution of catalyst states. However, current spatial resolution and sensitivity limitations hinder their efficacy. Advanced synchrotron-based infrared and Raman spectroscopies hold significant promise for capturing more precise surface changes, though their application in electrocatalytic $CO_2RR$ remains limited.

Integrating these advanced characterizations with ML modeling and DFT calculations can bridge the understanding of reaction center evolution, electrolyte-ion effects, and catalytic performance. Such integration would provide a more comprehensive framework for designing catalysts tailored for efficient $C_{2+}$ product formation. However, theoretical modeling still faces substantial challenges, particularly in accurately describing the electrolyte-electrode interface, where solvent effects must be considered.

For practical applications, the stability of catalysts at industrially relevant current densities remains a significant obstacle. A persistent discrepancy exists between performance in half-cell studies and full-cell tests. Bridging this gap requires systematic investigations to better replicate industrial conditions in lab-scale experiments. Despite recent progress, achieving Faradaic efficiencies exceeding 80% for $C_{2+}$ products remains rare. Furthermore, $C_{2+}$ products are often produced as mixtures with electrolytes, $H_2$, and residual $CO_2$, necessitating costly separation processes. These economic and technical barriers hinder the practical implementation of current electrocatalytic $CO_2RR$ technologies. In addition, on the anode side of the reaction, the oxygen evolution reaction (OER) remains kinetically sluggish, consuming nearly 90% of the total energy input.[186] Innovative strategies, such as coupling $CO_2RR$ with organic oxidation reactions or the chlor-$CO_2$ reaction, could reduce energy consumption while producing valuable co-products.[187,188] Novel electrolyzer designs, including $H_2$-$CO_2$ and metal-$CO_2$ batteries, offer dual functions as electrolyzers and energy storage devices, further enhancing efficiency.

### 5.3. Prospective

The discussions above underscore key avenues for advancing electrocatalytic $CO_2RR$, including coupled reactions, novel electrolyzer designs, comprehensive catalyst characterization, and computational modeling. Efficient production of $C_{2+}$ products hinges on the proper design of catalysts, incorporating not only C−C coupling mechanisms and associated engineering strategies but also considerations of other critical factors such as electrode/electrolyte interfaces, transport phenomena, catalyst stability, and electrolyte/solvent effects. A holistic approach from both mechanistic and system-level perspectives is essential.

Besides, the catalytic stability remains a persistent challenge, as performance in half-cell tests often deviates significantly from that in full-cell devices. For example, N-doped carbon and M−N−C (M: metal) catalysts, widely regarded as stable during electrocatalytic reactions, face destabilization in fuel cell devices due to the N protonation. Understanding the mechanisms of catalyst destabilization is critical to developing effective stabilization protocols. Currently, catalyst deactivation frequently arises from changes in structure, morphology, or oxidation states during operation. Emerging strategies focus on dynamic in situ reconstruction of active sites to restore or enhance activity. However, in situ reconstruction remains poorly understood, as it is influenced by multiple factors. Further studies are needed to elucidate the reconfiguration mechanisms and to establish robust structure-electrochemical performance relationships.

As for the catalyst materials, SACs and 2D materials offer promising platforms for atomic-level investigations of electrocatalytic $CO_2$RR mechanisms. SACs, in particular, have demonstrated reversible in situ reconstruction from isolated single-atom sites to clusters, which can act as active centers for $C_{2+}$ product formation. Innovations such as single-atom alloys and tandem SACs leverage synergistic interactions between neighboring atoms to facilitate C−C coupling. To promote the C−C coupling, high-density SACs with closely spaced active sites are anticipated, especially when supported on metastable 2D materials.[38] These ultrathin 2D platforms provide uniform electronic structures, high surface areas, and unique metastable forms with rich dangling bond states near the Fermi level. These features enable strong anchoring of catalysts, ranging from SACs to controlled metallic layers, enhancing catalytic performance. For example, Cu-based catalysts with amorphous or metastable surfaces could benefit from these principles, enabling precise control over active site structures. Compared to complex HEAs, SACs supported on 2D materials offer a more tractable platform for theoretical modeling and computational predictions. This facilitates a deeper understanding of the mechanisms governing $C_{2+}$ product formation and aids in the rational design of next-generation catalysts.

Building on the key factors involved in C−C coupling, the integration of advanced characterization techniques, in situ monitoring, and computational insights is anticipated to drive the development of efficient and reliable catalysts. These efforts will enable the selective and scalable production of $C_{2+}$ products through electrocatalytic $CO_2$RR. We hope the insights and strategies discussed in this report will contribute to a deeper understanding of C−C coupling mechanisms and catalyst engineering. By guiding the design and fabrication of electrocatalysts, these discussions aim to accelerate the transition of electrocatalytic $CO_2$RR technologies from fundamental research to practical applications.

## Acknowledgements

We gratefully acknowledge support from National Natural Science Foundation of China (52172228), and the Natural Science Foundation of Fujian Province (2024J01475 and 2023J05127). This report was proposed and convinced by Liangxu Lin. All other co-authors provided assistance, constructive suggestions on the writing and presentation.

**Conflict of Interest**

The authors declare no conflict of interest.

# TOC

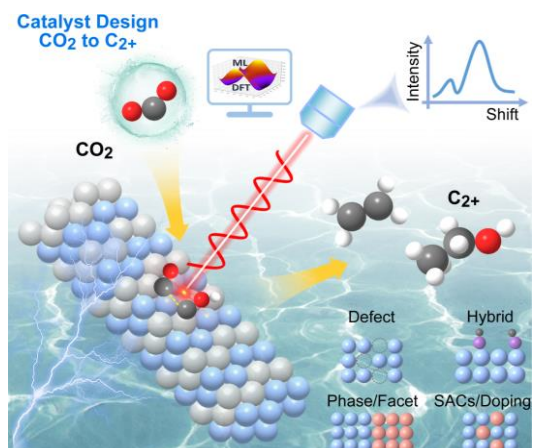

This report discusses recent advances in the selective production of $C_{2+}$ products via electrocatalytic $CO_2RR$ based on the underlying mechanisms, key intermediates and factors of C–C coupling. Insights and strategies aim to integrate advanced technologies and support the development of efficient catalysts for transitioning $CO_2RR$ from research to practical applications.

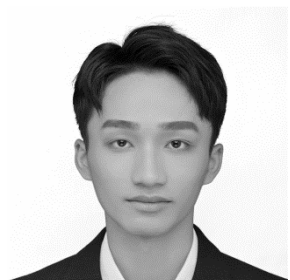

**Guangyuan Liang** is an MSc student at Fujian Normal University. He received his BSc degree from the Guangxi Normal University in 2022. His research focuses on the electrocatalytic technique for $CO_2$ reduction reaction.

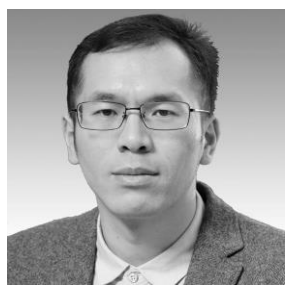

**Hongfang Du** is currently an Associate Professor at Fujian Normal University. He received his PhD degree in Clean Energy Science from Southwest University. He then worked as a research assistant at Nanyang Technological University and moved to Northwestern Polytechnical University as an assistant professor. His research interests focus on flexible electronics and clean energy conversion systems, including electrochemical water splitting, nitrogen reduction, carbon dioxide reduction.

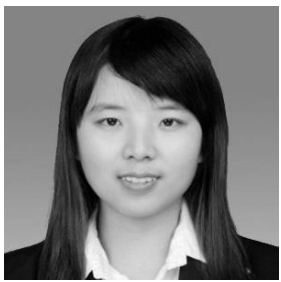

**Dandan Cui** received her Ph.D. degree from Beihang University in 2020. She is currently as a Lecturer of the Department of Physics at Beihang University. Her research interests focus on design and synthesis of 2D materials for energy conversion.

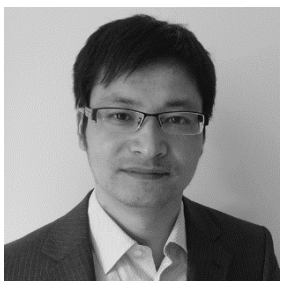

**Liangxu Lin** is currently a Full Professor at Fujian Normal University. He completed his PhD in Engineering Materials at The University of Sheffield at the end of 2013. After two postdoc periods at the University of Exeter, he joined Wuhan University of Science and Technology as a Professor, and the University of Wollongong as a Vice-Chancellor Fellow. His research focuses on two-dimensional nanomaterials for electrochemical energy storage/conversion, catalysis and materials interfaces.